\documentclass[]{aa}  

\usepackage{txfonts}
\usepackage[a4paper,
			breaklinks,
			unicode,
			colorlinks,
			linktocpage,
			pdftex,
			citecolor=blue,
			linkcolor=blue,
			urlcolor=blue,
			backref,
			pagebackref=false,
			bookmarks,
			bookmarksnumbered
			]{hyperref}

\hypersetup{
	pdftitle=The boundary layer in compact binaries,
	pdfauthor=Marius Hertfelder,
  	pdfcreator={LaTeX},
 	pdfproducer={LaTeX}
}

\begin{document}

\title{The boundary layer in compact binaries}

\author{Marius Hertfelder\thanks{\email{marius.hertfelder@gmail.com}}\inst{1}
	\and
		Wilhelm Kley\inst{1}
	\and
		Valery Suleimanov\inst{2,3}
	\and
		Klaus Werner\inst{2}
	}

   \institute{Institut für Astronomie und Astrophysik, Abt.~Computational Physics, Auf der Morgenstelle 10, 72076 Tübingen, Germany
		\and
		Institut für Astronomie und Astrophysik, Abt.~Astronomie, Sand 1, 72076 Tübingen, Germany
		\and
		Kazan (Volga region) Federal University, Kremlevskaya 18, 420008 Kazan, Russia
    }

\date{Received 27 August 2013 / Accepted 14 October 2013}

\abstract
	{
Disk accretion onto stars leads to the formation of a boundary layer (BL) near the stellar surface where the disk makes contact with the star.
Although a large fraction of the total luminosity of the system originates from this tiny layer connecting the accretion disk and the
accreting object, its structure has not been fully understood yet.
	}
	{
It is the aim of this work to obtain more insight into the BL around the white dwarf in compact binary systems. 
There are still many uncertainties concerning the extent and temperature of the BL and the rotation rate of the
white dwarf.
	}
	{
We perform numerical hydrodynamical simulations, where the
problem is treated in a one-dimensional, radial approximation (slim disk).
The turbulence is described by the $\alpha$ parameter viscosity.
We include both cooling from the disk surfaces and radial radiation transport. The radiation energy is treated in a one-temperature approximation.
	}
	{
For a given $\dot{M}$ our results show a strong dependence on the stellar mass and rotation rate.
The midplane and the effective temperature rise considerably with increasing stellar mass or decreasing stellar rotation rate.
Our simulations also show that the radiation energy and pressure are indeed important in the BL.
However, some models show a low optical depth in the BL,
making it necessary to find a better representation for optically thin regions.
	}
	{
The combination of a high mass and a small radius, characteristic of white dwarfs, can lead to an enormous energy release in the BL, provided the WD rotates slowly. Since the radial extent of BLs is typically very small (about $0.02$ to $0.05R_\ast$), this leads to surface temperatures of a few hundred thousand Kelvin. All of our models showed subsonic infall velocities with Mach numbers of $<0.4$ at most. 
	}


\maketitle

\titlerunning{The boundary layer in compact binaries}
\authorrunning{M.~Hertfelder et al.}

\section{Introduction}

Accretion of matter on compact objects via a disk is a ubiquitous phenomenon in astrophysics and can be observed in a variety of systems such as protostars,
close binary systems, and active galactic nuclei. 
Since accretion disks appear very frequently in astrophysical situations, they have been studied extensively and several
analytical models of stationary disks assuming axial symmetry and neglecting the vertical direction (the equations have been vertically integrated) have been constructed
\citep[e.g.][]{1973A&A....24..337S,1974MNRAS.168..603L,1981ARA&A..19..137P,1982SSRv...32..379V}. The vertical structure was investigated afterwards with the help of
these models \citep[e.g.][]{1982A&A...106...34M}.

While about one half of the accretion energy is released over almost the whole radial extent of the accretion disk, the other half is
still stored in terms of kinetic energy of the gas near the surface of the central object. Only when the flow reaches the surface of the star is it slowed down from
the Keplerian rotation rate 
to the rotation of the star, which will usually be much slower. This tiny
region with a radial extent of less than one percent of the stellar radius where the accretion disk connects to the central object is called the boundary layer
(BL). Since a great deal of energy is released in a spatially restricted area, the BL can get very hot and might account for the observed soft and hard X-ray and UV
emission in several cataclysmic variable systems \citep[e.g.][]{1981MNRAS.196....1C,1981ApJ...245..609C,1984MNRAS.206..879C}. It is therefore of great importance when it
comes to understanding star-disk systems. The first models of the BL were stationary calculations or estimates of timescales that were made under various
assumptions by different authors \citep{1974MNRAS.168..603L,1977MNRAS.178..195P,1977AcA....27..235T,1981AcA....31..267T,1979MNRAS.187..777P,1983A&A...126..146R}.
However, because of the simplification made there have been debates concerning the consistency of these models with the observations
\citep[e.g.][]{1982ApJ...262L..53F}. The first evolutionary calculations were performed soon afterwards
\citep{1986MNRAS.221..279R,1987A&A...172..124K,1989A&A...208...98K,1989A&A...222..141K,1991A&A...247...95K,1995MNRAS.275.1093G}, and for the first time allowed non-stationary phenomena like instabilities to be investigated. Among the first analytical studies of the BL were the efforts of \citet{1992ApJ...399L.163B} and \mbox{\citet{1995MNRAS.272...71R}}, who dealt with the one-dimensional, stationary equations by using the method of matched asymptotic expansions (MAE).

Most of the one-dimensional BL models focus on the radial evolution of the variables, where an infinitesimal thin disk is connected directly to the star. The vertical velocity is
set to zero and the fast moving accretion flow is slowed down within the midplane of the disk and accreted by the star. The
deceleration of the gas in the BL is accomplished by a mechanism that is not yet fully understood \citep[cf.][]{2012ApJ...751...48P,2012ApJ...760...22B}. 
Here, we assume a viscous medium and use the classical $\alpha$-parametrisation from 
\citet{1973A&A....24..337S}. The radiation emitted by the BL strongly depends on the mass accretion rate of the accreting object.
For rather low mass accretion rates ($\dot{M}\leq 10^{-10} M_{\sun}/\text{yr}$, \citealt{1987MNRAS.227...23W}), the BL is optically thin, reaching very high
temperatures ($\sim10^8\,\text{K}$) and emitting soft and hard X-rays in the case of cataclysmic variables (e.g. 
\citealt{2004RMxAC..20..244M,2003MNRAS.346.1231P,2005ApJ...626..396P},
\citealt{1984Natur.308..519K,1987Ap&SS.130..303S,1993Natur.362..820N,1999MNRAS.308..979P}). An optically thick BL, on
the contrary, shows a lower temperature of about $10^5\,\text{K}$ \citep{1977MNRAS.178..195P, 1986SvAL...12..117S, 1995ApJ...442..337P} and emits radiation that is mostly
thermalized and resembles a black body spectrum (see e.g. \citealt{1980MNRAS.190...87C,2004RMxAC..20..174M}).

It should be noted that in addition to the above-mentioned radial models, there are also one-dimensional models that investigate the vertical flow of the gas in the near
proximity of the star. Here, it is assumed that the gas spreads around the star because of the ram pressure in the BL. In contrast to the radial models,
the gas is not slowed down in the midplane of the disk, but rather on the whole surface of the star. This alternative model is called the spreading layer
\citep{1999AstL...25..269I,2010AstL...36..848I}. 
While this model was originally designed for the BL of a
neutron star, \citet{2004ApJ...616L.155P,2004ApJ...610..977P} adapted it for cataclysmic variables systems. 
The spreading layer concept was extended by \citet{2006MNRAS.369.2036S}, who included general relativity and
different chemical compositions of the accreted matter.

In contrast to these 1D models, a number of multidimensional studies of the BL have been performed, mostly under the assumption
of axial symmetry.
Full radiation hydrodynamical simulations of that kind were performed by \citet{1989A&A...208...98K,1989A&A...222..141K,1991A&A...247...95K}, but only very
few dynamical timescales could be followed in low spatial resolution.
In recent years, local BL models with higher resolution have been presented by \citet{2005ApJ...635L..69F,2006NewAR..50..509F} for the purely adiabatic cases.
Calculations including magnetic fields have been done in 2D by \citet{2003ApJ...589..397K} and in 3D by
\citet{2002MNRAS.330..895A}, though with only low resolution and short dynamical timescales. 
Additional 2D simulations were performed by \citet{1996ApJ...461..933K,1999ApJ...518..833K} for protostars,
by \citet{2008MNRAS.386.1038B} for neutron stars in LMXB,
or by \citet{2009ApJ...702.1536B}, who simulated the BL around a white dwarf and used a simplified
energy dissipation function without radiation transport. Three-dimensional magnetohydrodynamical simulations including magnetospheric accretion were
performed by \citet{2012MNRAS.421...63R}, but without the inclusion of radiation transport. Non-axial-symmetric phenomena were investigated only very recently by
\citet{2012ApJ...760...22B}, though only with an isothermal equation of state.

In this paper we focus on the BL structure around white dwarfs in compact binaries such as cataclysmic variable systems. 
Despite the existing multidimensional studies,
we opted for new one-dimensional models because of the moderate computational effort. These are made more realistic by including radiation transport in the radial direction and local cooling in the vertical direction. 
To allow variability studies, we solve the time-dependent equations. This extends the approach of \citet{1995ApJ...442..337P}, who solved the stationary equations (see Sect.~\ref{sec:comparison_pn95} for more details). Furthermore, we have implemented force and dissipation terms of the radiation field and a quasi-two-dimensional radiation transport to treat the radiation field consistently in our calculations, where the radiation pressure (energy) is comparable to the thermal pressure (energy). This work is the first step and will be used to expand the simulations to more dimensions. The results presented in this paper will be used to calculate synthetic, theoretical spectra and thereby considerably narrow the regimes for various parameters of binary systems, like the stellar rotation rate of the white dwarf. More detailed theoretical spectra and observational consequences will be presented in a subsequent paper.

The paper itself is organized as follows: In Sect.~2, an overview of the used equations and assumptions is given, and 
basic physics of the models is described. Section~3 is devoted to the numerical methods
that were utilized in order to solve the equations. 
In Sect.~4, the models are presented and discussed. We conclude with Sect.~5.

\section{Equations}\label{sec:equations}

In this section, we present the one-dimensional, vertically integrated Navier-Stokes equations used in the numerical code. Although one-dimensional BL calculations are certainly not sufficient to describe the structure of the BL, they are adequate to model the emitted energy, since the gas is slowed down in the midplane before it engulfs the star
\citep[e.g.][]{1989A&A...222..141K}.

\subsection{Vertical averaging}\label{sec:vertical_structure}

The 1D equations of motion are obtained through vertical integration of the 
Navier-Stokes equations over the $z$ coordinate. 
Assuming a Gaussian profile for the three-dimensional density $\rho$ in the vertical direction, 
the surface density is given by
\begin{equation}
	\Sigma = \int_{-\infty}^\infty \rho\,\text{d} z = \sqrt{2\pi}\rho_c(r)H~.
\end{equation}
Here, $\rho_c$ denotes the mass density in the midplane and $H$ is the pressure scale height, which is a measure of the vertical extent of the disk. If we
also assume hydrostatic balance and an isothermal equation of state in the vertical direction, the pressure scale height reads
\begin{equation}
	H = \frac{c_\text{s}}{\Omega_\text{K}}~,\label{eq:scale_height}
\end{equation}
where $c_\text{s}$ denotes the (isothermal) sound speed,
and $\Omega_\text{K}=\sqrt{GM_\ast/r^3}$ the Keplerian angular velocity ($G$ is the
gravitational constant and $M_\ast$ is the mass of the star).

In our 1D approximation, the vertical component of the velocity vector $\vec{u}$ is assumed to be negligible.
Formally, this can no longer be true in the regions where the gas leaves the midplane and spreads to
the poles of the star.

\subsection{Continuity equation (conservation of mass)}

The vertically integrated continuity equation in polar coordinates is then given by
\begin{equation}
	\frac{\partial\Sigma}{\partial t} + \frac{1}{r}\frac{\partial(r\Sigma u_r)}{\partial r} = 0~,\label{eq:cont_eq}
\end{equation}
where $u_r$ denotes the velocity component in the radial direction.

\subsection{Equations of motion (conservation of momentum)}

The vertically integrated equations of motion read
\begin{multline}\label{eq:radial_motion}
	\Sigma\left(\frac{\partial u_r}{\partial t} + u_r\frac{\partial u_r}{\partial r}-\frac{u_{\varphi}^2}{r}\right) = -\frac{\partial p}{\partial r} +
\frac{1}{r}\frac{\partial(r\sigma_{rr})}{\partial r} - \frac{1}{r}\sigma_{\varphi\varphi}\\ - \Sigma\frac{GM_\ast}{r^2} + \frac{\kappa_\text{R}\Sigma}{c}\vec{F}~,
\end{multline}
in the radial direction and
\begin{equation}\label{eq:azimuthal_motion}
	\Sigma\left(\frac{\partial u_{\varphi}}{\partial t} + u_r\frac{\partial u_{\varphi}}{\partial r} + \frac{u_ru_{\varphi}}{r}\right) =
\frac{1}{r^2}\frac{\partial(r^2\sigma_{r\varphi})}{\partial r}
\end{equation}
in the azimuthal direction. Here $p$ denotes the vertically integrated thermal pressure and $\kappa_\text{R}$ and $\vec{F}$ are the Rosseland mean opacity and the radiative flux, respectively, which we will describe in detail in the next section. The term $\frac{\kappa_\text{R}\Sigma}{c}\,\vec{F}$ acts as a radiative force on the material; 
$\sigma$ denotes the vertically integrated viscous stress tensor with the components $\sigma_{rr}, \sigma_{\varphi\varphi}$ and $\sigma_{r\varphi}$ \citep{1986MNRAS.220..593P},
where we assume a vanishing bulk viscosity.

\subsection{Energy equation (conservation of energy)}

Since the temperatures in the boundary layer region around white dwarfs are expected to be very hot even if they are optically thick \citep[the temperature is of the
order of $10^5\,\text{Kelvin}$,][]{1979MNRAS.187..777P,1981AcA....31..267T}, radiation pressure and radiation energy play an important role in our models and cannot be
ignored. Instead of simultaneously solving one equation for the gas energy and one for the radiation energy separately
(called the two-temperature approximation\footnote{
Although often used, this term might be misleading since the radiation energy cannot be described by any temperature if no LTE is assumed.
}), we
chose a different approach in order to speed up the calculations. 
In this so-called one-temperature radiation transport \citep[see e.g.][]{2010MNRAS.409.1297F},
we add the two equations for the gas and the radiation energy and obtain
\begin{equation}
	\rho\frac{\text{d}}{\text{d} t}\left(\varepsilon + \frac{E}{\rho}\right) = -P \nabla\vec{u} + \left[\sigma_{ij}-\mathcal{P}_{ij}\right]\nabla_i u_j -
\nabla\vec{F} \,,
\end{equation}
where $\varepsilon,E,P,\mathcal{P}$ and $\vec{F}$ denote the specific thermal energy of the gas, 
the radiative energy density, the 3D thermal pressure of the gas, the radiation pressure tensor, and the radiative flux, respectively. 

The assumption of a local thermodynamic equilibrium is justified for optically thick regions, as is the one-temperature approach \citep[see e.g.][]{2010A&A...511A..81K}.
Since we concentrate here on high mass accretion rates during outbursts, the BL will stay optically thick even for slow stellar rotation
rates. In the approximation of local thermal equilibrium (LTE), the radiation energy simplifies to read $E=aT^4$,
where $a$ is the radiation constant and $T$ is the gas temperature.

In addition to the one-temperature approximation, we use the flux-limited diffusion approximation (FLD)
\citep{1981ApJ...248..321L,1984JQSRT..31..149L}, which allows us not to consider an equation for the radiative momentum. The radiation flux is then set to
\begin{equation}
	\vec{F} = -\frac{c\lambda}{\kappa_\text{R}\rho}\nabla E = -\frac{\lambda c 4 aT^3}{\kappa_\text{R}\rho}\nabla T~,
\end{equation}
where $\lambda$ is a dimensionless number called the flux-limiter.
Here, we adopt the formulation by \citet{1981ApJ...248..321L},
\begin{align}
	\lambda & = \frac{1}{R}\left(\coth R-\frac{1}{R}\right)~,\\
	R & = \frac{|\nabla E|}{\kappa_\text{R}\Sigma E}~.
\end{align}

For the given flux-limiter, the corresponding approximation for the radiation pressure tensor reads \citep{1984JQSRT..31..149L}
\begin{equation}
	\mathcal{P}= \frac{E}{2}\left[(1-f_\text{Edd})\mathcal{I} + (3f_\text{Edd}-1)\,\vec{n}\vec{n}\right]~.\label{eq:pressure_tensor}
\end{equation}
Here, $\mathcal{I}$ is the identity tensor of rank 2, $\vec{n}=(\nabla E)/|\nabla E|$ is the unit vector parallel to the gradient of $E$, and the Eddington factor is given by
\begin{equation}
	f_\text{Edd}=\lambda + \lambda^2R^2~.
\end{equation}
This approximation reflects the correct behaviour of the radiation pressure tensor in the optically thick regime where it is isotropic and $E/3$, and in the optically thin regions where its absolute value parallel to $\nabla E$ is $E$. If one considers a purely isotropic radiative pressure tensor, a approximation in the flux-limited diffusion theory is given by $\mathcal{P}=\lambda E\mathcal{I}$ \citep{2011A&A...529A..35C}. Our simulations showed that the difference between Eq.~(\ref{eq:pressure_tensor}) and the purely isotropic approximation is in general very small.

Furthermore, if we assume that the radiation pressure tensor is diagonal (Eddington approximation) and use the relation between the specific thermal energy and temperature
$\varepsilon=c_\text{v}T$, the
energy equation in the one-temperature approximation becomes after a vertical integration
\begin{equation}\label{eq:one_temp_temp}
\begin{split}
 & \left[\Sigma c_v + 4aT^3\cdot\widetilde{H} \right] \left(\frac{\partial T}{\partial t} + u_r\frac{\partial T}{\partial r}\right) 
 		+ 4aT^4\cdot\frac{1}{r}\frac{\partial(ru_r)}{\partial r}  = \\
	&\quad -p\frac{1}{r}\frac{\partial(r u_r)}{\partial r} - \mathcal{P}_{ij}\nabla_i u_j \\
	&\quad + 2 \nu \Sigma \left[\left(\frac{\partial u_r}{\partial r}\right)^2 + \left(\frac{u_r}{r}\right)^2\right]
		   + \nu \Sigma \left(r\frac{\partial\Omega}{\partial r}\right)^2 -\frac{2}{3} \nu \Sigma\left(\frac{1}{r}\frac{\partial(r u_r)}{\partial r}\right)^2 \\
	&\quad - 2\sigma_\text{SB} T_\text{eff}^4
		   + \widetilde{H}\frac{1}{r}\frac{\partial}{\partial  r}\left[\frac{16\sigma_\text{SB}\lambda}{\kappa_\text{R}\Sigma}r\widetilde{H}T^3\frac{\partial
T}{\partial r}\right]~,	
\end{split}
\end{equation}
where $\sigma_\text{SB}$ is the Stefan-Boltzmann constant, $T_\text{eff}$ the effective temperature, $\kappa_\text{R}$ the Rosseland opacity, and
$\widetilde{H}=\sqrt{2\pi}H$. The second line describes the pressure work exerted by the thermal and radiative pressure. The third line contains viscous dissipation, where $\nu$ denotes the kinematic viscosity. The last line describes emission of radiation from the disk surface and diffusion of the radiative flux in the disk midplane. By this means we employ a quasi-2D radiation transport.

To close the set of equations, we need some constitutive relationships. For the equation of state, we use the ideal gas law. The vertically integrated pressure then reads
\begin{equation}
	p = \frac{\Sigma R_\text{G} T}{\mu}~,
\end{equation}
where $R_\text{G} = k_\text{B}/m_\text{H}$ with $k_\text{B}$ being the Boltzmann constant, $m_\text{H}$ the mass of hydrogen, and $\mu$ is the mean molecular weight. In order to take the radiation pressure effects into account, we have defined an effective sound speed \citep{2007ApJ...667..626K}
\begin{equation}
	c_\text{eff} = \sqrt{\frac{\gamma p + (4/9)\cdot E\cdot\widetilde{H}\cdot(1-e^{-\kappa_\text{R}\rho\Delta r}) }{\Sigma} }~.
\end{equation}
The factor $(1-e^{-\kappa_\text{R}\rho\Delta r})$ ($\Delta r$ is the width of a cell) is used to interpolate between the optically thick region, where the radiation pressure increases the effective sound speed since it contributes to the restoring force and optically thin regions, where radiation pressure plays no role.

For the opacity we use Kramer's law,
\begin{equation}
	\kappa = \kappa_0 (\rho/\text{g cm}^{-3}) (T/\text{K})^{-3.5}~,
\end{equation}
where $\kappa_0 = 5\times 10^{24}\,\text{cm}^2\,\text{g}^{-1}$. If the temperature is high enough for the gas to be fully ionized, we can assume a lower threshold for
the opacity
given by free electron scattering processes (Thomson scattering). The corresponding opacity has the constant value $\kappa_\text{Thomson} =
0.335\,\text{cm}^2\,\text{g}^{-1}$ (assuming a hydrogen mass fraction of $X=0.675$ for the gas composition) and is added to Kramer's opacity. Because we 
estimate the local cooling of the disk via a blackbody radiation of temperature $T_\text{eff}$, we need a relation that links the effective temperature to the
temperature in the midplane of the disk. Therefore, we employ the relation by \citet{1990ApJ...351..632H} which is a generalisation of the grey atmosphere
\citep[e.g.][]{1986rpa..book.....R} and approximates the optical depth in the vertical direction of the disk. The relations read \citep{1992SvAL...18..104S}
\begin{align}
	T^4 & = \tau_\text{eff}T_\text{eff}^4 \label{eq:t_eff}\\
	\tau_\text{eff} & = \frac{3}{8}\tau_\text{R} + \frac{\sqrt{3}}{4} + \frac{1}{4\tau_\text{P}}~,\label{eq:tau_eff}
\end{align}
where $\tau_\text{R}$ and $\tau_\text{P}$ ($\tau = \kappa\rho H=\frac{1}{2}\kappa\Sigma$) are the Rosseland and the Planck mean optical depth \citep[see also][]{2008A&A...487L...9K}.\label{sec:t_eff}

\subsection{Viscosity}

The mechanism that accounts for the angular momentum transport in the boundary layer region is still a matter of concern
\citep{1994MNRAS.268...29P,1994ApJ...431..359N,1994PASJ...46..289K,1995MNRAS.277..157G}. The most likely driving force for the anomalous
viscosity in accretion disks with magnetic Prandtl numbers of the order of unity
is the magneto-rotational instability \citep[e.g.][]{2008NewAR..51..814B} 
that gives rise to the onset of turbulence
\citep{1991ApJ...376..214B,1998RvMP...70....1B,2003ARA&A..41..555B} which acts like a genuine viscosity on macroscopic scales. However, this cannot be the case for the boundary
layer, since the magneto-rotational instability is effectively damped out for a increasing rotation profile $\Omega(r)$ \citep{1995MNRAS.277..157G,1996MNRAS.281L..21A}. For lack of a better representation, we assume that the angular momentum transport in the BL is managed by turbulence of some kind
\citep[cf.][]{1981ARA&A..19..137P}. In that case we can use the classic $\alpha$-prescription by \citet{1973A&A....24..337S}, which is a parametrisation for the stresses
caused by turbulence in an accretion disk and therefore is still valid for MRI unstable disks, provided that a viable value for the numerical parameter $\alpha$ is
given. This $\alpha$ ansatz, which is a frequently used expression for the disk viscosity is written as
\begin{equation}
	\nu = \alpha c_\text{s}H\,, \label{eq:alpha_viscosity} 
\end{equation}
where $c_s = \sqrt{p/\Sigma}$ is the isothermal sound speed. Unless stated otherwise, Eq.~(\ref{eq:alpha_viscosity}) was used to calculate the viscosity in our models.
In the BL, the radial pressure scale height becomes smaller than the vertical one. This is considered in the viscosity prescription by \citet{1986MNRAS.220..593P}, which reads
\begin{equation}
	\nu = \alpha c_\text{s}\left[\frac{1}{H^2} + \frac{(\text{d} p/\text{d} r)^2}{p^2}\right]^{-1/2}~.\label{eq:pn95_viscosity}
\end{equation}
We used Eq.~(\ref{eq:pn95_viscosity}) when we compared our calculations with that of \citet{1995ApJ...442..337P} in Sect.~\ref{sec:comparison_pn95}.

\section{Numerical methods}

\subsection{General remarks}

The partial differential equations, Eqs.~(\ref{eq:cont_eq}, \ref{eq:radial_motion}, \ref{eq:azimuthal_motion}) and (\ref{eq:one_temp_temp}), are discretized on a fixed Eulerian grid using the finite differences method and propagated in time using a
semi-implicit-explicit scheme. For this purpose a new framework, guided by the \texttt{ZEUS} code by \citet{1992ApJS...80..753S}, has been programmed completely
from scratch and tested extensively before it was used to perform the calculations. 
To ensure a formal second-order accuracy in time and space, we employed a staggered grid in space \citep[see e.g.][]{1979CoPhC..18..171T} and a multistep procedure for the time integration \citep[operator-splitting, e.g.][]{1984ApJS...55..211H}. The computational domain typically ranges from one to two stellar radii, $\mathcal{D}=\left[R_\ast,2R_\ast\right]$, and is divided into 512 logarithmically spaced grid cells.

\subsection{Implicit methods}

For some source terms, a special treatment for the time integration is necessary, because using a time-explicit scheme would constrain the time step considerably and
slow down the simulations. Hence those parts of the equations, namely the radial diffusion (radiation transport) and the viscous torques and forces, have to be solved
with an implicit scheme that does not limit the time step if we are looking for a stationary solution. In contrast to an explicit time integration, the equations are now
solved assuming the value of the physical quantity at the new time step $n+1$. This leads to a system of linear equations, which can be written as a matrix-vector
multiplication in a space of dimension $N$, where $N$ is the number of active grid cells. However, this system can become very large if we are using a large number of
cells. Fortunately, in our one-dimensional case the use of this implicit method only leads to a tridiagonal matrix, and the equations can be solved easily.

\subsection{The time step}

One undesirable feature of explicit numerical methods in hydrodynamics is the limitation of the largest possible time step \citep{Courant28}. 
It must be limited with respect to the characteristics of the system according to the relation known as the CFL condition,
\begin{equation}
	\Delta t = f_\text{C}\cdot\text{min}_j\left[\frac{\Delta r_j}{\left|u_j\right| + c_{\text{s},j} }\right]~,\label{eq:cfl}
\end{equation}
where $\Delta r_j$ is the extent of the cell $j$, $u_j$ its velocity, and $c_{\text{s},j}$ the sound speed. 
The minimum of all active cells is taken. 
Typically, we use a Courant factor $f_\text{C}=0.8$ in Eq.~(\ref{eq:cfl}).

\subsection{Boundary and initial conditions}\label{sec:boundary_initial_conditions}

The modelling of the boundary layer surrounding a star is a classical boundary value problem, meaning that the solution we try to obtain must satisfy not only the
partial differential equations but also the boundary conditions because of the finite space domain. In mathematical terms those
conditions are given as either Dirichlet or Neumann boundary conditions where either a value or the normal derivative of a variable is specified. 

Physically speaking, at the outer boundary of our computational domain, we have to allow for an incoming mass flux that corresponds to the accretion of matter.
Additionally, we require the angular velocity to be Keplerian $\Omega(r_\text{out})=\sqrt{GM/r_\text{out}^3}$ and consider a pressure correction. At the inner boundary,
where the stellar surface is located, we set the angular velocity of the gas to the stellar rotation rate $\Omega(r_\text{in}) = \Omega_\ast$, i.e. we impose a no-slip
boundary condition in $\varphi$-direction. Instead, the radial velocity is not set to zero, but rather to a small but finite fraction of the Keplerian velocity,
\begin{equation}
\label{eq:ur_in}
u_r(r_\text{in}) = \mathcal{F}\sqrt{GM_\ast/r_\text{in}} \,,
\end{equation}
in order to enable a mass flux out of the domain.
For $\mathcal{F}$ we choose a value much smaller than one, typically $10^{-5}$. 
This open inner boundary is necessary to avoid the accumulation of mass and allow for accretion onto the star.
The value of  $\mathcal{F}$ determines how much of the stellar interior is taken into account in the simulation.
The extreme case of  $\mathcal{F}=0$ would imply that the whole star is considered which does not make sense in this 
type of simulation. Our choice of  $\mathcal{F}$ ensures that a sufficient part of the stellar envelope is
contained in the domain but not too much, otherwise the temperatures become too high and slow down the simulations.
For testing purposes we have varied  $\mathcal{F}$ and do not find any differences in the results despite a small shift in
radius.

The stellar radiation is taken care of by including the flux $F_\ast =
\sigma_\text{SB}T_\text{eff}^4$ in the radiative diffusion routine as an inner boundary condition. 
Because the temperature at $r_\text{in}$ is inside the star, it is not known a priori and cannot be specified.
Hence, a simple zero gradient condition for the temperature is assumed in all other routines that require
a temperature.
Previous studies suggest that the thermal boundary condition at the star is of 
great importance for the BL \citep{1995MNRAS.272...71R,1995MNRAS.275.1093G}.
We would like to point out that the influence of the incoming flux is virtually nil, as test calculations have shown. 
Apart from the Dirichlet conditions that have been shown above, for most other physical variables we
imposed zero gradient boundary conditions, which means that the normal derivative at the boundary equals zero (Neumann type).

Initially, we have to prescribe values for all variables. 
Here, it is vital to ensure that the initial conditions are compatible with the boundary conditions and are physically reasonable. 
We have found that a reliable set of initial conditions is given by the \citet{1973A&A....24..337S} disk solution \citep[see e.g.][for a compact representation]{frank2002accretion},
even though they are not completely  consistent with all boundary conditions. 
To avoid any problems in this context, we have interpolated both regions smoothly. For a given stellar mass $M_\ast$, radius $R_\ast$, temperature $T_\ast$, and rotation rate $\omega_\ast$, the solution will be given by the mass inflow rate $\dot{M}$ and the viscosity parameter $\alpha$.

\subsection{Model parameters}

In this paper, we focus on the boundary layer around the white dwarf in cataclysmic variable systems, and investigate its structure. We computed several models,
where we vary the mass of the white dwarf and its rotation rate. We ran the simulations with the following three masses, $M=0.8
M_{\sun}, 1.0M_{\sun}$, and $1.2M_{\sun}$, which are typical white dwarf masses in cataclysmic variables. 
Another important parameter that also determines the entire amount of energy released
in the accretion disk and BL is the radius of the star. Here we used the mass-radius relation from
\citet{1972ApJ...175..417N}. We imposed an effective temperature of $T_\ast = 50\,000\,\text{K}$ that is, for example, consistent with the estimates of \citet{2010ApJ...716L.157S} for SS Cyg, where a mass accretion rate of $\dot{M} = 1.51\times10^{-8} M_{\sun}/\text{yr}$ was assumed.
Since we are interested in the thermodynamics of the BL, the rotation rate of the
white dwarf is an important parameter. It determines how much energy is dissipated in the BL region and therefore has a major influence on the temperature. We simulated a variety of
models where the white dwarf is non-rotating ($\Omega_\ast = 0$), fast rotating ($\Omega_\ast=0.8\Omega_\text{K}$), and in between. 
We took $\alpha=0.01$ for the viscosity parameter throughout, which is probably too small a choice for the disk. In the BL, however, the viscosity is supposed to be far smaller than in the disk. Test calculations with $\alpha=0.1$ showed no major structural differences compared to the models presented here, except for the inflow velocity that reaches the sonic point for models with high mass and low rotation rates. If we use Eq.~(\ref{eq:pn95_viscosity}) instead of the classical $\alpha$ ansatz for the viscosity, the sonic point is hit only for greater values of $\alpha$. To avoid unphysical, supersonic infall velocities, it is possible to include a causality preserving factor \citep[e.g.][]{1992ApJ...394..261N}.

\section{Results}

\subsection{The $1.0 M_{\sun}$ model}

First, we will describe the basic properties of the boundary layer for our standard model of a one solar mass white dwarf.
In doing this, we are going to emphasise the dependence of the structure and thermodynamics on the rotation rate of the central star, a parameter whose exact value is still unclear in many systems. We usually
show five different stellar rotation rates (a fraction of $\{0.0,0.2,0.4,0.6,0.8\}$ of the Keplerian rotation rate at the stellar surface) and use the abbreviation
$\omega := \Omega_\ast / \Omega_\text{K}(R_\ast)$.

\subsubsection{Dynamic structure of the disk}

\begin{figure}[t]
  \begin{center}
    \includegraphics[scale=.45]{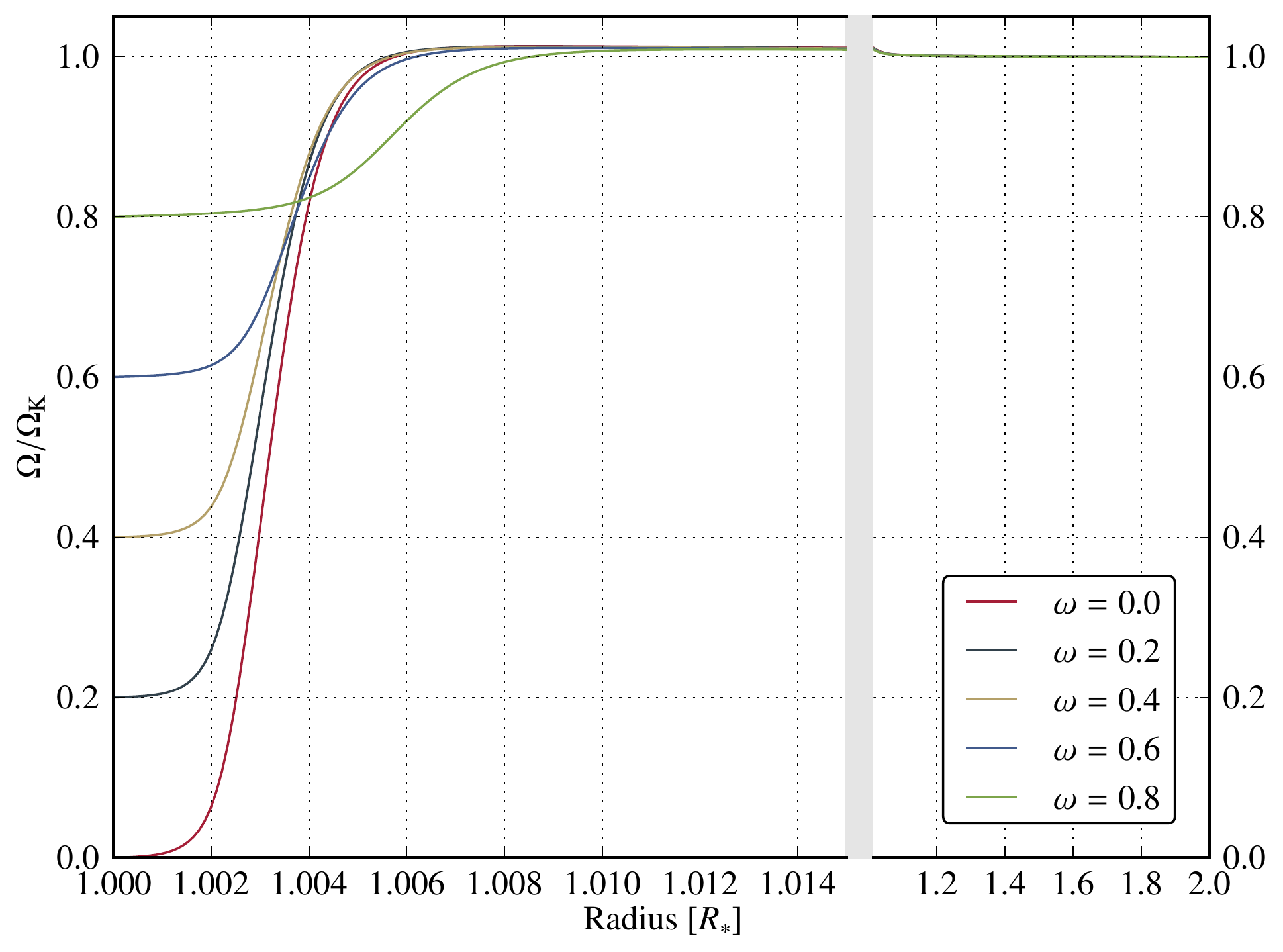}
    \caption{\label{fig:10_Om}Angular velocity $\Omega=u_{\varphi}/r$ in terms of the Keplerian angular velocity for a central star with mass $M_\ast=1.0 M_\sun$ and five different stellar rotation rates $\Omega_\ast$, denoted by
$\omega=\Omega_\ast/\Omega_\text{K}(R_\ast)$. The plot depicts two regions separated by a light grey bar that differ in the radial scaling and enable us to show both
the rapid variation in the BL and the constant overall trend.}
  \end{center}
\end{figure}

Figure \ref{fig:10_Om} shows the angular rotation rate of the gas in units of the Keplerian rotation rate, $\Omega_\text{K}=\sqrt{GM_\ast/r^3}$. We can clearly see that
outside the boundary layer, for $r/R_\ast\gtrsim 1.2$, the gas rotates with the Keplerian angular velocity. 
When moving farther inwards, the
gas rotates slightly super-Keplerian in order to compensate the large inward-pointing pressure gradient that is present in this region. Not until the gas is at a distance of less than a percent of the stellar radius does its angular velocity decrease to connect smoothly to the
stellar rotation rate. The more the angular velocity differs from the Keplerian value, the more the gas loses radial stabilisation via the centrifugal force. This is,
however, compensated by a large, now outward-pointing pressure gradient since the star has a much higher temperature and density and so the pressure is much higher
than in the BL.
Although the connection between the angular velocity of the gas and the star does not involve discontinuities, the angular velocity strongly changes in a very small
region. 
The width of
the BL is defined as the distance from the star (here $R_*$) to the point where the radial derivative of $\Omega(r)$ vanishes.
We note that this point does not in general coincide with the point where
$\partial/\partial r (\Omega/\Omega_\text{K})=0$. Table \ref{tab:width_BL} gives an overview of the width of the BL for different stellar rotation rates. We also note that $R_\ast$ is not  defined unambiguously owing to the continuous transition to the star, and the observed width
may depend on the value of $\mathcal{F}$, see Eq.~(\ref{eq:ur_in}).
With an increasing stellar rotation rate, the width of the BL decreases up to $\omega=0.2$ at first.
After that, however, it gets wider again. In general, the width of the
BL is governed by the viscosity, which in turn depends on the temperature and the surface density for the case of an $\alpha$-prescription. An increasing
temperature and surface density leads to a broadening of the BL region. A look at Fig.~\ref{fig:10_Tc}, which shows the temperature in the midplane of the disk,
indicates, that the temperature in the disk decreases with increasing stellar rotation rate. This behavior is in line with our expectations, since a faster moving star
means less friction and therefore less heating. As a consequence of this overall temperature variation with $\omega$, the faster the rotation of the star, the smaller the
width of the BL; however, the surface density rises with increasing $\omega$, as can be seen in the inset in Fig.~\ref{fig:10_Sigma}, because of
an increasingly inefficient mass transport through the disk (more mass can accumulate in the disk). This is the reason for the turnaround of the trend at $\omega=0.2$,
when the BL starts to become broader again. 
Another interesting feature is the comparatively large width of the BL for $\omega=0.8$ that stands out both in Table \ref{tab:width_BL} and Fig.~\ref{fig:10_Om}, and does not match the shape of the other models. This effect is caused by a slightly different temperature evolution, as can be seen in the inset in Fig.~\ref{fig:10_Tc}. Here, as a result of reduced friction and hence less energy release, the green line (representing $\omega=0.8$) is missing a peak
(compared to the other curves) and is wider than most other temperatures over a small region. 

\begin{table}
	\caption{\label{tab:width_BL}Width of the boundary layer for $M_\ast=1.0 M_\sun$ and different stellar rotation rates
$\omega=\Omega_\ast/\Omega_\text{K}(R_\ast)$. By definition, the BL ranges from the
surface of the star to the point where $\partial\Omega(r)/\partial r =0$, i.e. where it has a maximum. Additionally, the absolute value of $u_\varphi=\Omega\cdot r$ at $r_\text{max}$ is given as
$u_\varphi^\text{max}$.}
	\begin{center}
		\begin{tabular}{c c c c}
		\hline\hline
		$\omega$ & $\Delta r$ $[R_\ast]$ & $u_\varphi^\text{max}$ $[\text{cm}\,\text{s}^{-1}]$ & $\beta = \frac{ \Omega_\text{max} }{ \Omega_\text{K}(r_\text{max}) }$\\\hline
		$0.0$ & $0.0074$ & $493\times 10^{6}$ & $1.013$\\
		$0.2$ & $0.0072$ & $493\times 10^{6}$ & $1.012$\\
		$0.4$ & $0.0074$ & $492\times 10^{6}$ & $1.012$\\
		$0.6$ & $0.0079$ & $492\times 10^{6}$ & $1.011$\\
		$0.8$ & $0.0104$ & $490\times 10^{6}$ & $1.008$\\
		\hline
		\end{tabular}
	\end{center}
\end{table}

\begin{figure}[t]
  \begin{center}
    \includegraphics[scale=.45]{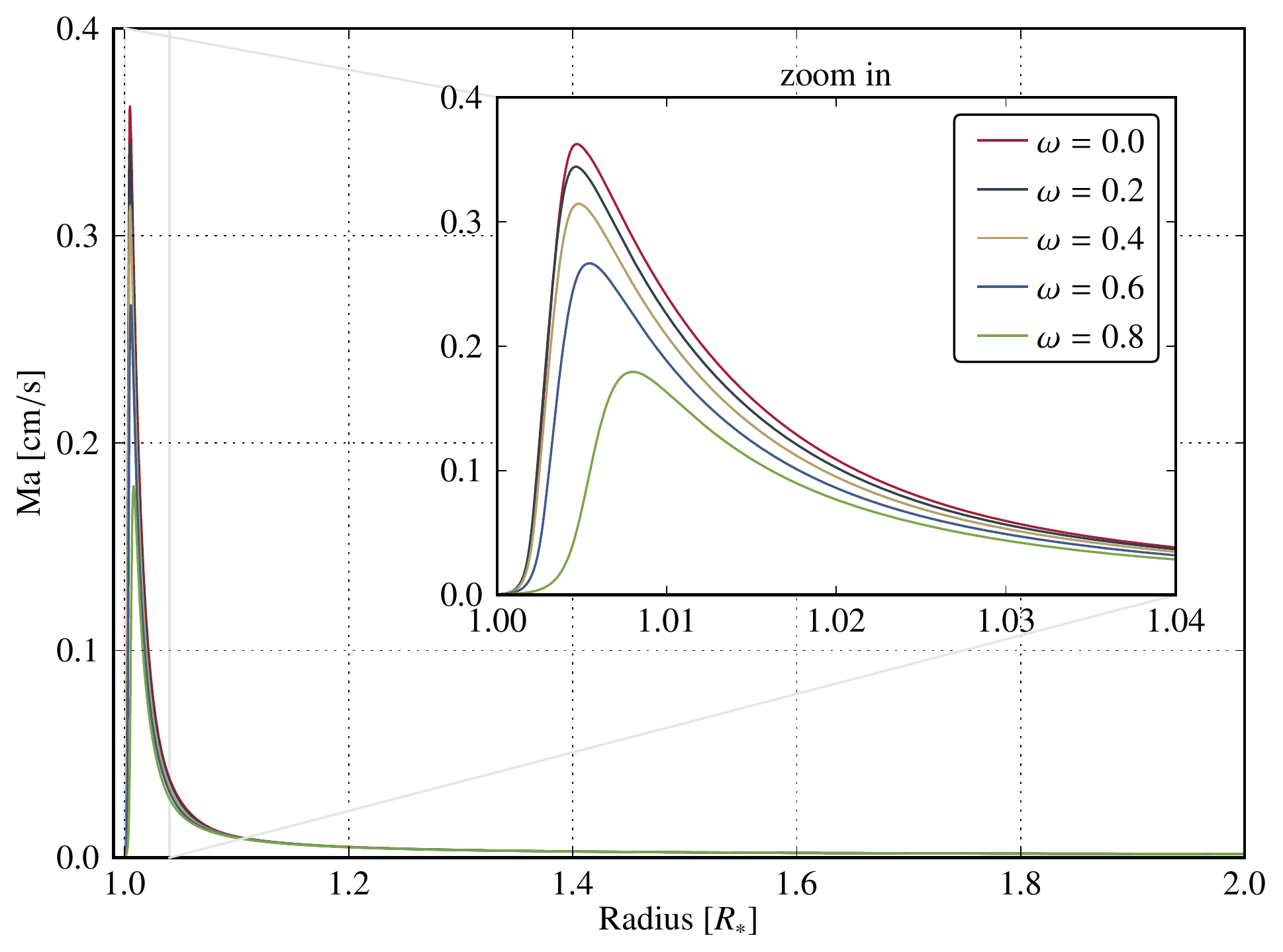}
    \caption{\label{fig:10_vrad}The plot shows the radial Mach number of the gas for a stellar mass of $1.0M_\sun$. The Mach number is defined as a quotient of radial
velocity and the sound speed $\text{Ma}=|u_r|/c_\text{s}$. The different colours correspond to the different stellar rotation rates
$\omega=\Omega_\ast/\Omega_\text{K}(R_\ast)$. The smaller box inside the graph is a zoom in of the inner edge; the light grey frame denotes the zoom area.}
  \end{center}
\end{figure}

The Mach number of the gas, which is defined as the quotient of the radial velocity and the speed of sound $\text{Ma}=|u_r|/c_\text{s}$ is shown in Fig.~\ref{fig:10_vrad}. Since the radial velocity is negative throughout the computational domain, in principle the Mach number outlines the
velocity of the radially inward-falling material. While the gas is moving inwards with a rather low velocity over most parts of the disk, there is a distinct maximum of the
radial velocity in the BL.
What drives the material to move inwards more rapidly in the BL is the loss of angular momentum caused by friction in the disk.
Hence the radial velocity is
increasing strongly in the BL region, as can be observed in the inset in Fig.~\ref{fig:10_vrad}. After peaking in the BL, the radial
velocity decreases again as the gas approaches the surface of the star. Here, the gas slows down as it settles onto the atmosphere of the star, where the individual
layers are stabilized by the pressure (hydrostatic balance). The inflow velocity of the gas depends on how much angular momentum can be removed
from the material, which in turn is dependent on the strength of the shearing. Accordingly, the faster the stellar rotation rate, the weaker is the shearing in the disk
and hence the radial velocity should decrease with increasing stellar rotation rate in the BL. We can observe this trend in Fig.~\ref{fig:10_vrad} where we can also see that the radial velocity is clearly subsonic throughout the
computational domain and especially in the BL. Thus there are no problems concerning causality \citep[e.g.][]{1977MNRAS.178..195P} in our simulations because of the small value of $\alpha$.

\subsubsection{Thermal structure of the disk}

In the previous section, we looked closely at the dynamical structure of the disk and the boundary layer, which is determined by the radial and azimuthal
velocities. From an observational point of view, however, we are much more interested in the thermodynamics of the disk, since the quantity that we can actually observe,
the emitted radiation, depends on the temperature; therefore, we will now explore the surface density and the temperature structure of our models.

\begin{figure}[t]
  \begin{center}
    \includegraphics[scale=.45]{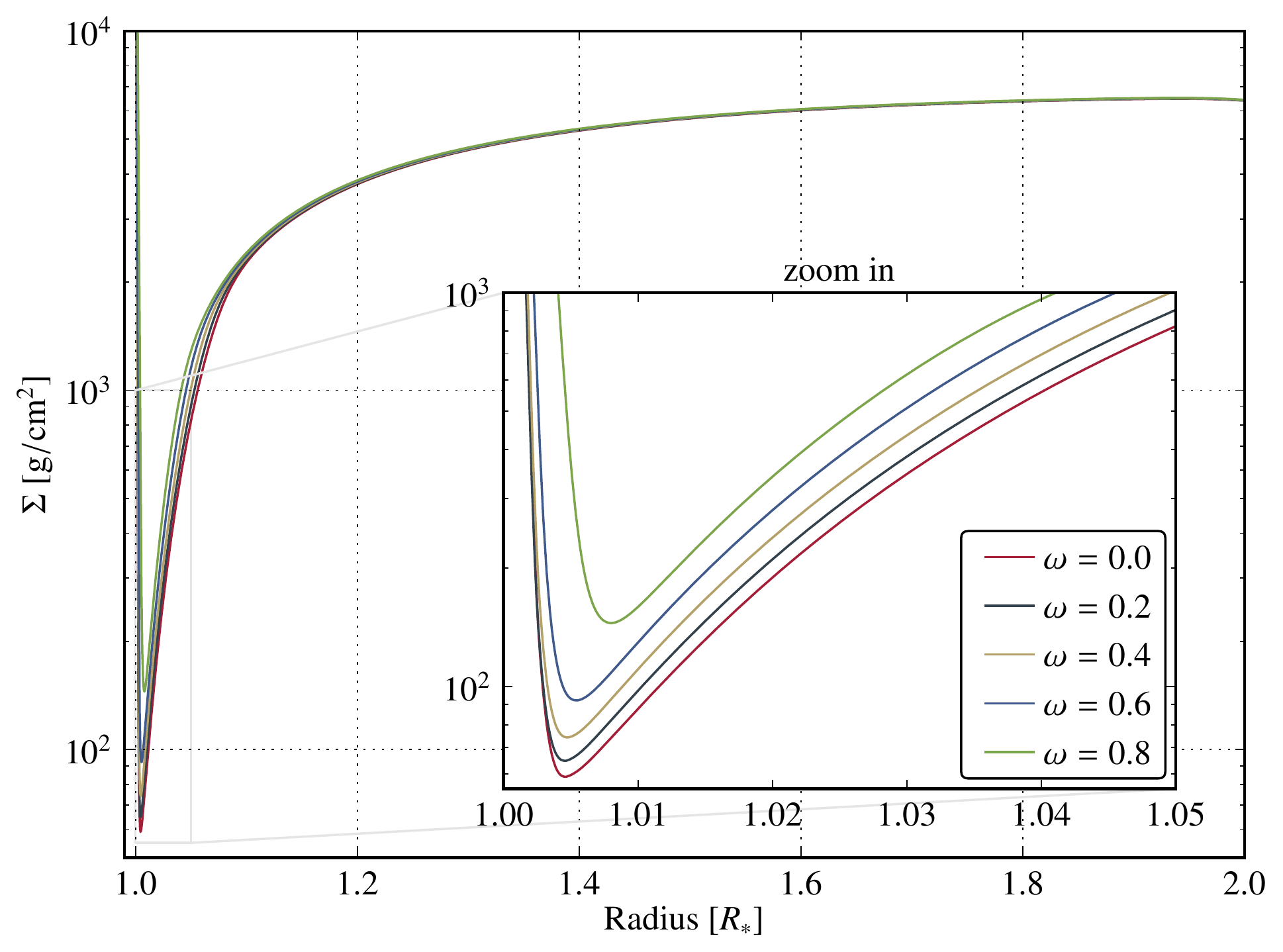}
    \caption{\label{fig:10_Sigma}Surface density $\Sigma\sim\rho H$ (log-scale) of the disk and a stellar mass of $1.0M_\sun$ for five different stellar rotation rates
$\omega=\Omega_\ast/\Omega_\text{K}(R_\ast)$. The inset is a zoom in of the inner edge; the light
grey frame denotes the zoom area.}
  \end{center}
\end{figure}

The surface density $\Sigma$ for the $1.0M_\sun$ model is shown in Fig.~\ref{fig:10_Sigma}. We can observe a heavy depletion of material in the BL. The
surface density drops by approximately two orders of magnitude compared to the weakly varying value in the disk. If we go farther in, the surface density rises rapidly
since we encounter the surface of the star which has a density that is orders of magnitudes higher than in the disk. The reason for the strong decrease of the
surface density in the BL is the aforementioned increase in inflow velocity. Since for the equilibrium state the mass accretion rate $\dot{M}$ is constant
throughout the disk (see also Fig.~\ref{fig:10_mdot} below), a local increase in inflow velocity leads to a reduced density at this point. The BL resembles a
bottleneck with a higher velocity and a lower density. In the disk the distinction between models
with different values of $\omega$ is very small; it is more pronounced in the BL.
The models
with a slower rotating star have a smaller surface density in the BL that is accounted for by the greater inflow velocity.

\begin{figure}[t]
  \begin{center}
    \includegraphics[scale=.45]{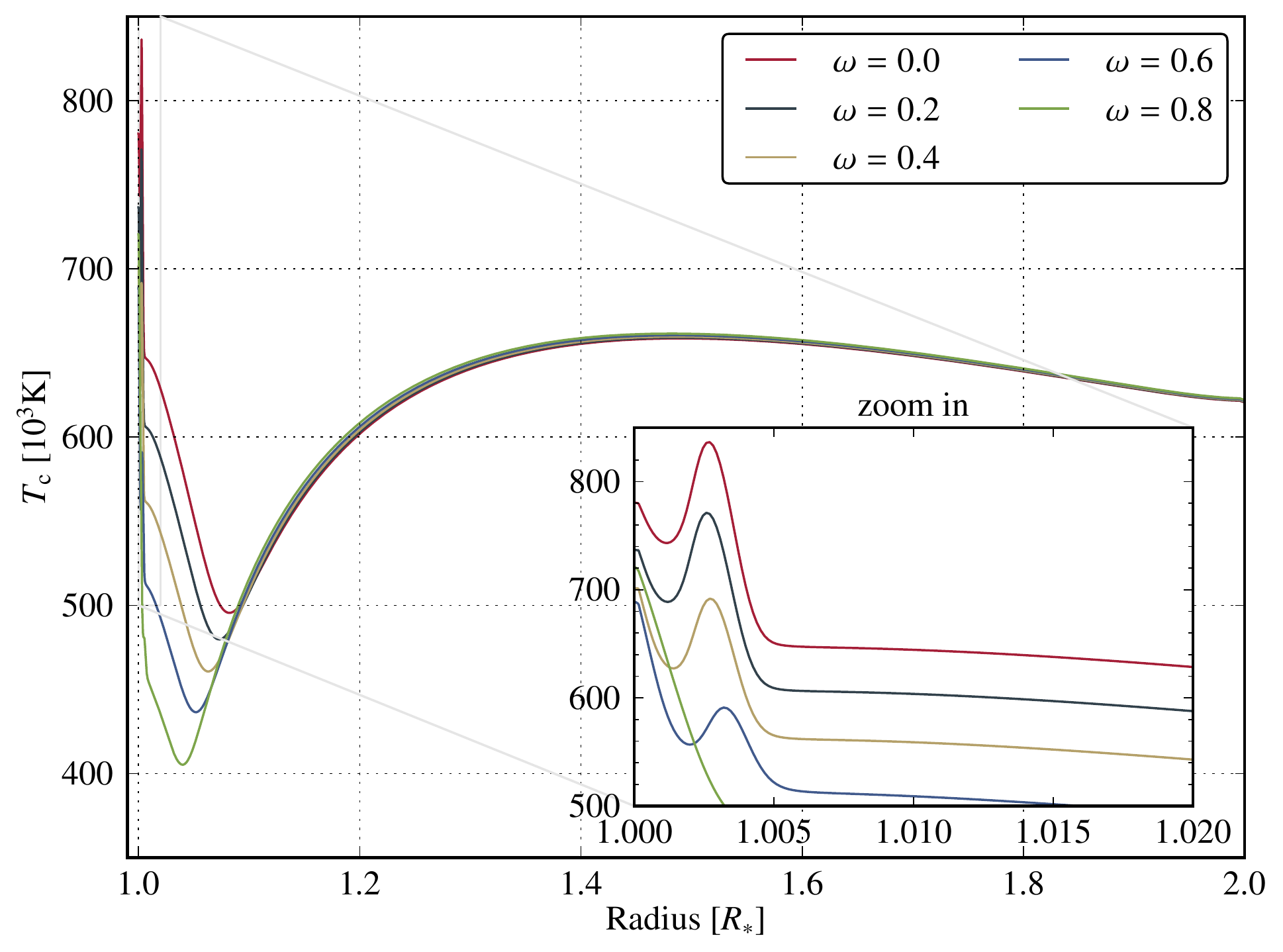}
    \caption{\label{fig:10_Tc}The graph shows the temperature $T_\text{c}$ in the midplane of the disk, i.e. $z=0$, for $M_\ast=1.0M_\sun$. The five models displayed
differ in the stellar rotation rate $\omega=\Omega_\ast/\Omega_\text{K}(R_\ast)$. The inset is a zoom in of the inner edge; the light
grey frame denotes the zoom area. The small peak in the inset is a clear indication of a hot BL.}
  \end{center}
\end{figure}

Even more interesting than the density structure is the temperature of the disk. In our one-dimensional models, we distinguish between the temperature in the midplane of
the disk $T_\text{c}$ and the effective or surface temperature, which is close to the disk temperature at $\tau\approx 1$. Figure
\ref{fig:10_Tc} shows the midplane temperature for $M_\ast=1.0M_\sun$ and five different stellar rotation rates. The local heat production is proportional to the
square of the shear. Hence in the disk the temperature decreases with increasing radius $r$. 
In the BL, $\Omega$ reaches its maximum and the shear in the gas vanishes; therefore, there is a location in the disk where the viscous dissipation vanishes and the local heating is approximately zero (apart from pressure work). For that reason, the temperature must drop in the region where $\partial\Omega/\partial r = 0$, so in or near the BL. The minimum of
$T_\text{c}$ does, however, not coincide with the maximum of $\Omega$, since heat is transported radially through the disk via advection and radiation. Furthermore, the
cooling of the disk through emission of radiation depends on the local optical depth and hence is different at each location. On the other hand, after the zero-gradient
point of $\Omega$, the shearing is very strong where the angular velocity connects to the stellar rotation rate. As a consequence, the temperature again rises, whereby
the considerably declined surface density compensates the strong shearing to some extent. The little peak in $T_\text{c}$, which is located very close to the star in the
BL, is a result of the heat production in the BL (see the inset in Fig.~\ref{fig:10_Tc}) and is a clear indication of a hot BL. The influence of the
stellar rotation rate on the temperature is twofold. In the disk, the temperature is slightly hotter for faster stellar rotation rates because of the larger
surface density. This trend is reversed in the area of the BL, where a smaller rotation rate causes more shear and supersedes the influence of the surface
density. A special case is obviously given for very fast rotating stars, as can be seen in Fig.~\ref{fig:10_Tc} for $\omega=0.8$, 
where the BL is broader and is missing a peak in the midplane temperature. The overall temperature regime of $400\,000$ to $840\,000$ Kelvin is very hot.

\begin{figure}[t]
  \begin{center}
    \includegraphics[scale=.45]{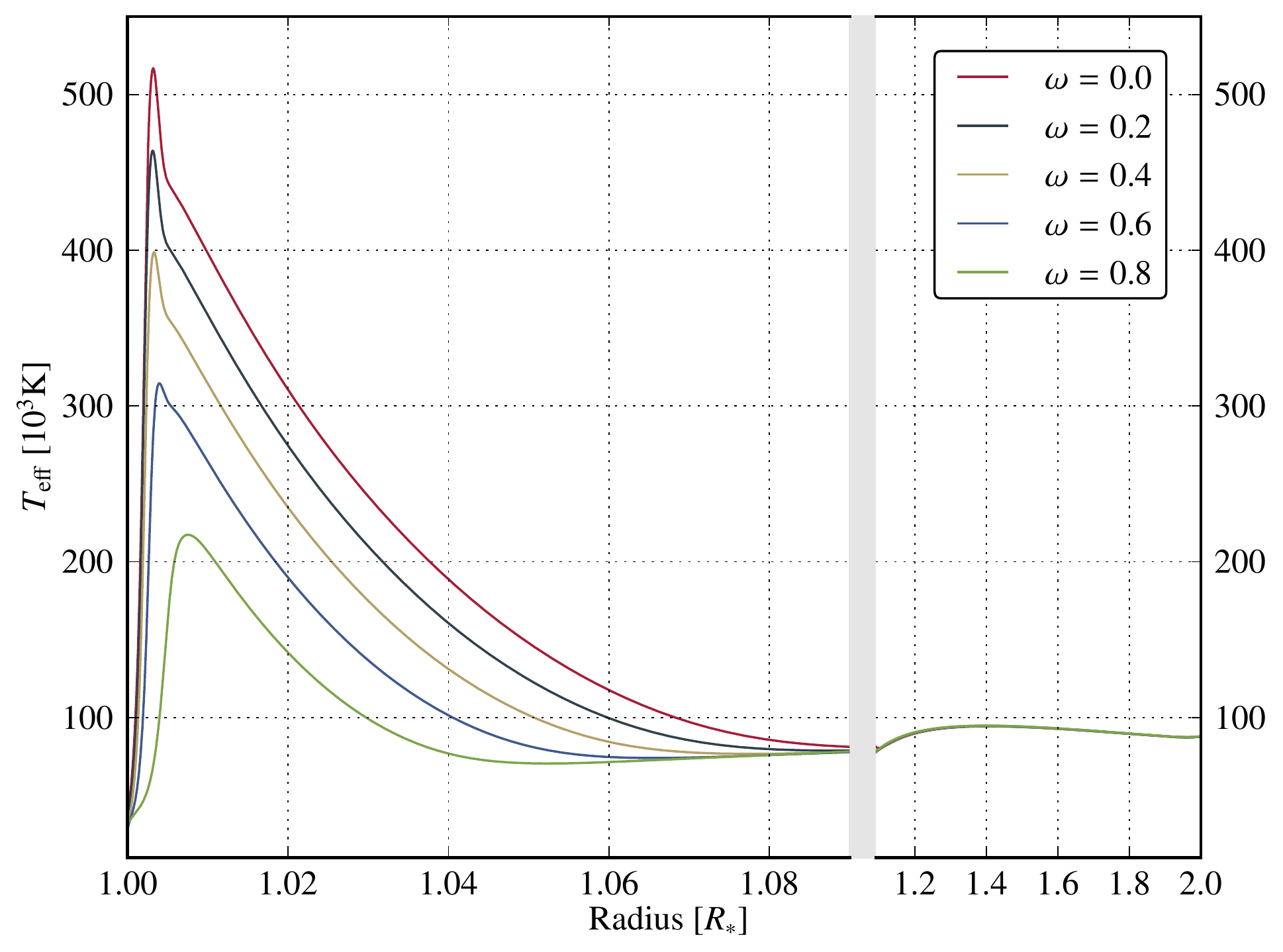}
    \caption{\label{fig:10_teff}Effective Temperature $T_\text{eff}$ of the disk, calculated according to Eq.~(\ref{eq:t_eff}). The plot shows five different models, all
of which have the same stellar mass of $1.0M_\sun$, but different stellar rotation rates $\omega=\Omega_\ast/\Omega_\text{K}(R_\ast)$. The plot depicts two regions
separated by a light grey bar that differ in the radial scaling and enable us to show both the narrow but very distinct peak of the effective temperature in the
BL and the nearly constant behaviour in the disk all in one graph.}
  \end{center}
\end{figure}

While the temperature shown in Fig.~\ref{fig:10_Tc} corresponds to the temperature in the midplane of the disk, we are particularly interested in the emergent
spectrum from the disk, which -- in the local thermodynamic equilibrium (LTE) approximation -- is determined by the temperature on the surface, see Eq.~(\ref{eq:t_eff}). The effective temperature $T_\text{eff}$ is shown in
Fig.~\ref{fig:10_teff} for the $M_\ast=1.0M_\sun$ case. For outer parts of the simulation domain, $T_\text{eff}$ changes only slightly and the difference in temperature
between various choices of $\omega$ is hardly noticeable. The reason for the almost constant behavior of the effective temperature in the disk is the strong increase in
the BL that somewhat masks the variation in the disk. While the midplane temperature changes at most by a factor of 2 over the whole simulation area, the
effective temperature changes considerably more, by a factor of 4-5. This is caused by a strong drop in the optical depth in the BL by several
orders of magnitude due to the drop of the surface density. While, generally, the accretion disk is optically thick, under certain circumstances, the BL can become optically thin, since the dilution of matter is severe in this region. 
While the shear and the energy production are confined to a small region called the
dynamical boundary layer, the release of the produced energy occurs over a slightly wider area, called the thermal boundary layer \citep{1995MNRAS.272...71R,1995ApJ...442..337P}. Therefore, the peak in
Fig.~\ref{fig:10_teff} is less intense than the one in Fig.~\ref{fig:10_Tc}, but instead it is wider because of the radial diffusion. According to what has
already been said above, the magnitude of the peak of $T_\text{eff}$ in the BL depends on the stellar rotation rate to the effect that a slower rotating star
causes more shear in the BL, hence a higher energy dissipation and eventually a higher effective temperature. 

\begin{figure}[t]
  \begin{center}
    \includegraphics[scale=.45]{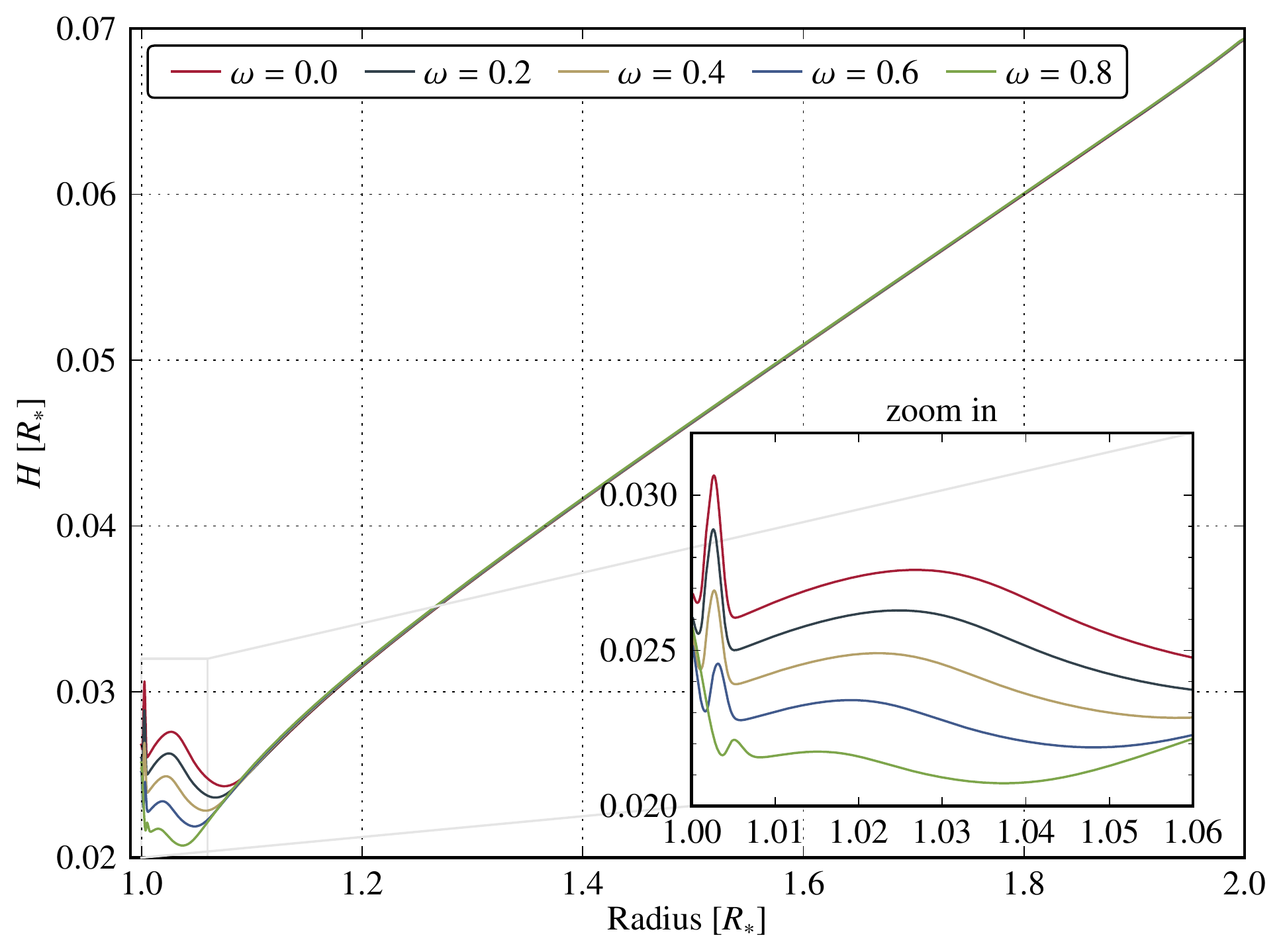}
    \caption{\label{fig:10_H}The graph shows the scale height $H$ (see Eq.~\ref{eq:scale_height}) of the disk for $M_\ast=1.0M_\sun$ and five different stellar rotation
rates $\omega=\Omega_\ast/\Omega_\text{K}(R_\ast)$. The disk is obviously very thin and not flared. In the BL, it is puffed up to some extent, which is,
however, not unexpected. The inset is a zoom in of the inner edge; the light grey frame denotes the zoom area.}
  \end{center}
\end{figure}

Finally, we investigate the vertical extent of the disk. According to Sec.~\ref{sec:vertical_structure}, a measure for the height is the variable $H$, which is
the point where the density drops by a factor of $1/\sqrt{e}$ compared to the midplane density. The scale height $H(r)$ is depicted in Fig.~\ref{fig:10_H} for the $1.0
M_\sun$ model. The disk in the system is, according to our simulations, rather thin with an aspect ratio $h=H/r$ of roughly $h=0.029$ (arithmetic mean) and has a slope of $\text{d}H/\text{d}r\approx0.05$. In the region of the BL, the disk height increases
noticeably because of the strong growth of temperature in the BL that causes a high pressure which puffs up the disk in the innermost region. Figure \ref{fig:10_H}
confirms the picture of the BL being a bottleneck that we introduced earlier. When approaching the BL from the outside, the disk first
starts to get thinner and thinner until it suddenly grows in height. In the disk the vertical extent is nearly identical for
different $\omega$, but in the BL it clearly depends on the stellar rotation rate. The faster the star spins, the less puffed out the BL is. We have
already pointed out, that the temperature depends on the stellar rotation rate. Since the pressure depends on the temperature, the scale height in the BL changes
with different stellar rotation rates.

\subsection{Dependence on the stellar mass}

After having discussed the basic properties of the boundary layer using the example of a solar mass white dwarf, we now focus on the dependence of the
BL on the stellar mass. First, we will again examine the dynamical structure of the BL (see Table \ref{tab:width_BL_comp} and Fig.~\ref{fig:10_Om}\footnote{
Additional figures can be found at: \url{http://www.tat.physik.uni-tuebingen.de/~hertfelder/BL2013}.
}). The shape of the angular velocity profile is identical for each of the three stellar masses. There is,
however, a difference in the width of the BL. The higher the stellar mass, the less broad is the BL. Table \ref{tab:width_BL_comp} shows the width of the
BL for the three different stellar masses and two different stellar rotation rates, along with the maximum value of $\Omega$ for $\omega=0.0$. Here it becomes
more obvious that the width of the BL strongly depends on the mass of the central star. The absolute value of the angular velocity also depends on the stellar mass and
varies significantly. The reason for the decreasing width of the BL is the mass radius relation of white dwarfs. The more massive the star is, the smaller the stellar radius $R_\ast$ is. This causes a nonlinear variation of gravity and so affects the width of the BL.
The Mach number stays more or less the same and the inflow occurs subsonically throughout all models.

\begin{table}
\caption{\label{tab:width_BL_comp}Width of the boundary layer for three different stellar masses and the stellar rotation rates of $\omega=0.0$ and $\omega=0.8$.
By definition, the BL ranges from the surface of the star to the point where $\partial\Omega(r)/\partial r =0$, i.e. where it has a maximum.}
	\begin{center}
		\begin{tabular}{c c c c}
		\hline\hline
		model & $\Delta r$ $[R_\ast]$ & $\Delta r$ $[R_\ast]$ & $u_{\varphi}^\text{max}$ $[\text{cm}\ \text{s}^{-1}]$\\
		&$\omega=0.0$&$\omega=0.8$&$\omega=0.0$\\\hline
		$M_\ast = 0.8M_\sun$ & $0.0086$ & $0.0134$ & $391\times 10^{6}$\\
		$M_\ast = 1.0M_\sun$ & $0.0074$ & $0.0104$ & $493\times 10^{6}$\\
		$M_\ast = 1.2M_\sun$ & $0.0066$ & $0.0084$ & $638\times 10^{6}$\\
		\hline
		\end{tabular}
	\end{center}
\end{table}

\begin{table}
\caption{\label{tab:comp_temps}Illustration of the midplane and effective temperature for the models with $0.8$ and $1.2M_\sun$. The table states the maximum temperatures both in the disk and in the BL for each stellar rotation rate.}
	\begin{center}
		\begin{tabular}{l l r r}
		\hline\hline
		\multicolumn{4}{c}{Midplane Temperature $T_\text{c}$}\\
		\hline
		&&\multicolumn{2}{c}{peak [$\text{K}$]}\\
		$M_\ast$ [$M_\sun$]	& $\Omega_\ast$ [$\Omega_\text{K}$]						& \multicolumn{1}{c}{BL} 	& \multicolumn{1}{c}{disk} 	\\
		\hline
		$0.8$	& $0.0$	& 	$640\,000$					&	$520\,000$							\\
				& $0.2$	& 	$590\,000$					&	$520\,000$							\\
				& $0.4$	& 	$520\,000$					&	$520\,000$							\\
				& $0.6$	& 	$450\,000$					&	$520\,000$							\\
				& $0.8$	& 	no peak						&	$520\,000$							\\
		\hline
		$1.2$	& $0.0$	& 	$1\,140\,000$				&	$900\,000$							\\
				& $0.2$	& 	$1\,060\,000$				&	$900\,000$							\\
				& $0.4$	& 	$950\,000$					&	$900\,000$							\\
				& $0.6$	& 	$820\,000$					&	$900\,000$							\\
				& $0.8$	& 	$650\,000$					&	$900\,000$							\\
		\hline\hline
		\multicolumn{4}{c}{Effective Temperature $T_\text{eff}$}\\
		\hline
		&&\multicolumn{2}{c}{peak [$\text{K}$]}\\
		$M_\ast$ [$M_\sun$]	& $\Omega_\ast$ [$\Omega_\text{K}$]						& \multicolumn{1}{c}{BL} 	& \multicolumn{1}{c}{disk} 	\\
		\hline
		$0.8$	& $0.0$	& 	$400\,000$					&	$64\,000$					\\
				& $0.2$	& 	$360\,000$					&	$64\,000$					\\
				& $0.4$	& 	$300\,000$					&	$64\,000$					\\
				& $0.6$	& 	$235\,000$					&	$64\,000$					\\
				& $0.8$	& 	$165\,000$					&	$64\,000$					\\
		\hline
		$1.2$	& $0.0$	& 	$690\,000$					&	$125\,000$							\\
				& $0.2$	& 	$625\,000$					&	$125\,000$							\\
				& $0.4$	& 	$540\,000$					&	$125\,000$							\\
				& $0.6$	& 	$440\,000$					&	$125\,000$							\\
				& $0.8$	& 	$300\,000$					&	$125\,000$							\\
		\hline
		\end{tabular}
	\end{center}
\end{table}

The dynamical structure is affected by the mass of the central star, and also the thermodynamics of the BL. To gain more insight into this
dependence, we first consider the surface density $\Sigma$ for the different models.  
The general trend of the $0.8$ and $1.2M_\sun$ models is closely related to that shown in Fig.~\ref{fig:10_Sigma}. We note, however, 
that the more massive the central star is, the higher the surface density in the BL and in the disk is, for the same reason as for the decrease of
the BL width, namely a lower effective viscosity. If the viscosity becomes smaller, the transport of matter through the disk is not as effective as it is for a higher
viscosity because the shear viscosity drives the accretion of matter. Thus, more matter can accumulate in the disk and the surface density rises. Again, the
minima of $\Sigma$ are located at different radii, corresponding to different widths of the BL. Apart from these features, the evolution of the surface density looks
identical for different stellar masses. The midplane temperature, on the other hand, shows considerable differences in magnitude between the three models. As 
Table~\ref{tab:comp_temps} shows, the $1.2M_\sun$ model is hotter than the $0.8M_\sun$ by a factor of $1.7$ throughout the BL and the disk. This
is due to the higher surface density in the high stellar mass model. Since the viscous dissipation is proportional to the surface density, an increasing $\Sigma$ ensures
an increase in the midplane temperature. This effect prevails, although the viscosity is, as we have seen, smaller than in the low mass models. 

\begin{figure}[t]
  \begin{center}
    \includegraphics[scale=.45]{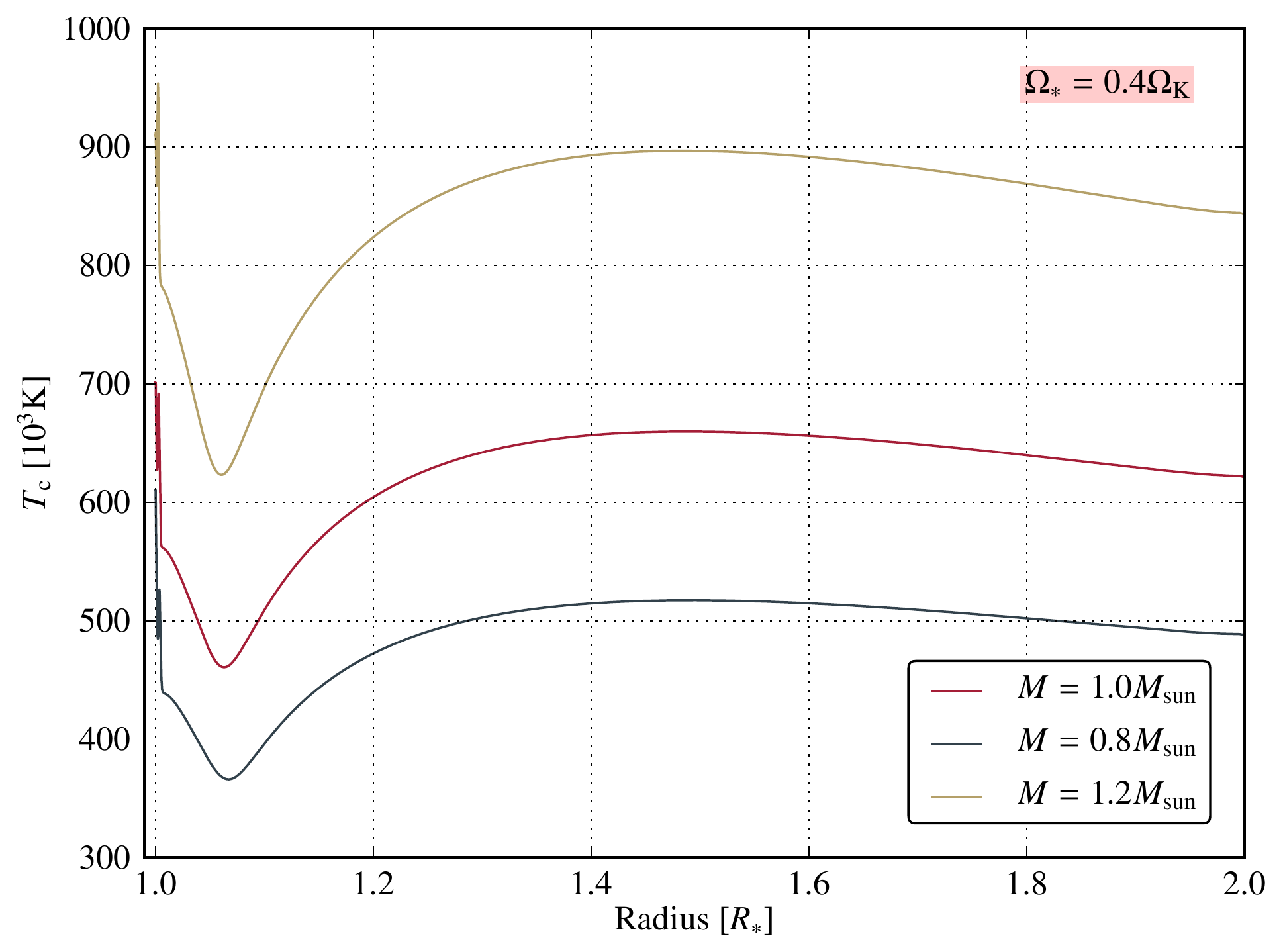}
    \caption{\label{fig:comp_tc}Comparison of the temperature in the disk midplane $T_\text{c}$ of the three different models with stellar masses of $1.0,
0.8$, and $1.2 M_\sun$. The stellar rotation rate $\Omega$ has been chosen to amount to $0.4\Omega_\text{K}$, as an illustrative case. It is clearly observable that with
increasing stellar mass the temperature in the disk also rises.}
  \end{center}
\end{figure}

A direct
comparison is shown in Fig.~\ref{fig:comp_tc}, where we have plotted $T_\text{c}$ for the three different stellar masses and $\omega=0.4$ in one single diagram. This
makes it clear how the midplane temperature increases overall with increasing stellar mass.

\begin{figure}[t]
  \begin{center}
    \includegraphics[scale=.45]{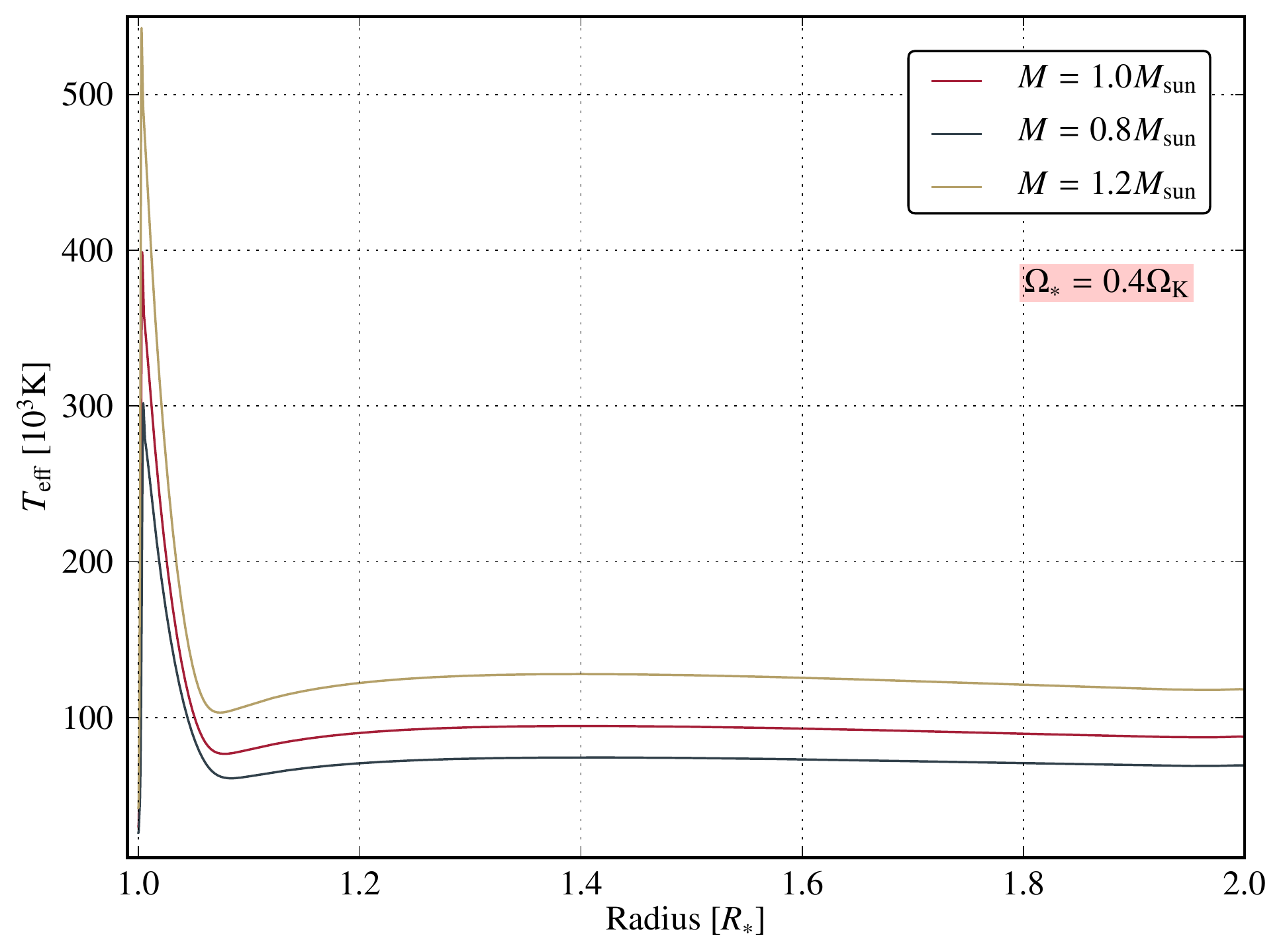}
    \caption{\label{fig:comp_teff}Analogous to Fig.~\ref{fig:comp_tc}, we compare the effective temperatures of the three different stellar masses ($1.0,0.8$, and
$1.2M_\sun$) in this plot. Again, $\Omega=0.4\Omega_\text{K}$ has been chosen as an example of the stellar rotation rate. We note that the effective temperature rises with
increasing stellar mass.}
  \end{center}
\end{figure}

Table~\ref{tab:comp_temps} also describes the effective temperature $T_\text{eff}$ for the three
different masses, whose general trend is given by Fig.~\ref{fig:10_teff}. As we have expected, the more massive the central star is, the higher the effective temperature of the BL (and also the disk) is and the
harder the radiation emerging from the BL is, because of the ratio of stellar mass and radius, which exclusively determines the total amount of accretion energy that
is released in the disk and the BL. The total luminosity that can be extracted from the process of accretion on a body with mass $M$ and radius $R$ is given
by
\begin{equation}
	L_\text{acc} = \frac{GM\dot{M}}{R}~,\label{eq:acc_en}
\end{equation}
where $\dot{M}$ is the mass accretion rate and $G$ the gravitational constant. The luminosity of the BL is at most one half of $L_\text{acc}$. Therefore, the greater the relation $M_\ast/R_\ast$, the higher the effective temperature. The inverse mass-radius-relation of white dwarfs enhances $L_\text{acc}$ even more. We can also see that the width of the thermal BL follows the trend of the dynamical BL. With increasing stellar mass, the width of the peak
of $T_\text{eff}$ decreases. Again, we have plotted all three stellar masses in one diagram (Fig.~\ref{fig:comp_teff}) that emphasizes the points
mentioned above. To give an overview of the radiation energy that corresponds to these temperatures, we focus on two models that are located at the opposite edges of the parameter space.
The $\omega=0.0, 1.2M_\sun$ model peaks at nearly $700\,000$ Kelvin. This corresponds to a black body radiation spectrum where the maximum of the distribution is located at a photon energy of $300\,\text{eV}$. The other model, where the parameters are given by $\omega=0.8, 0.8M_\sun$, peaks at an effective temperature of about $170\,000$ Kelvin. This corresponds to a photon energy of $73\,\text{eV}$ at the maximum of the black body spectrum. The effective temperature, or in other words, the spectrum emerging from the BL, is obviously a good way to try to determine the actual mass of the white dwarf in cataclysmic variable systems.

Finally, we want to have a look at the vertical structure of the disk. 
The scale height $H$ of the different models again follows the trend of Fig.~\ref{fig:10_H}. The aspect ratio of the $0.8M_\sun$ model is $H/r \approx 0.032$, while the model with $1.2M_\sun$ is slightly thinner with $H/r\approx0.026$. The solar
mass model lies in between with $H/r\approx0.029$, as has been said earlier. We can therefore deduce that with increasing stellar mass, the vertical extent of the
accretion disk diminishes. This also holds true for the BL, even though the midplane temperature is much higher in the high mass model and accordingly the
pressure will try to puff out the BL with much more force. However, the gravitational field of the central star exerts a force against the pressure and is strong
enough to outrange it. The disks are still not flared for all three stellar masses.

\subsection{Mass accretion rate and angular momentum flux}

\begin{figure}[t]
  \begin{center}
    \includegraphics[scale=.45]{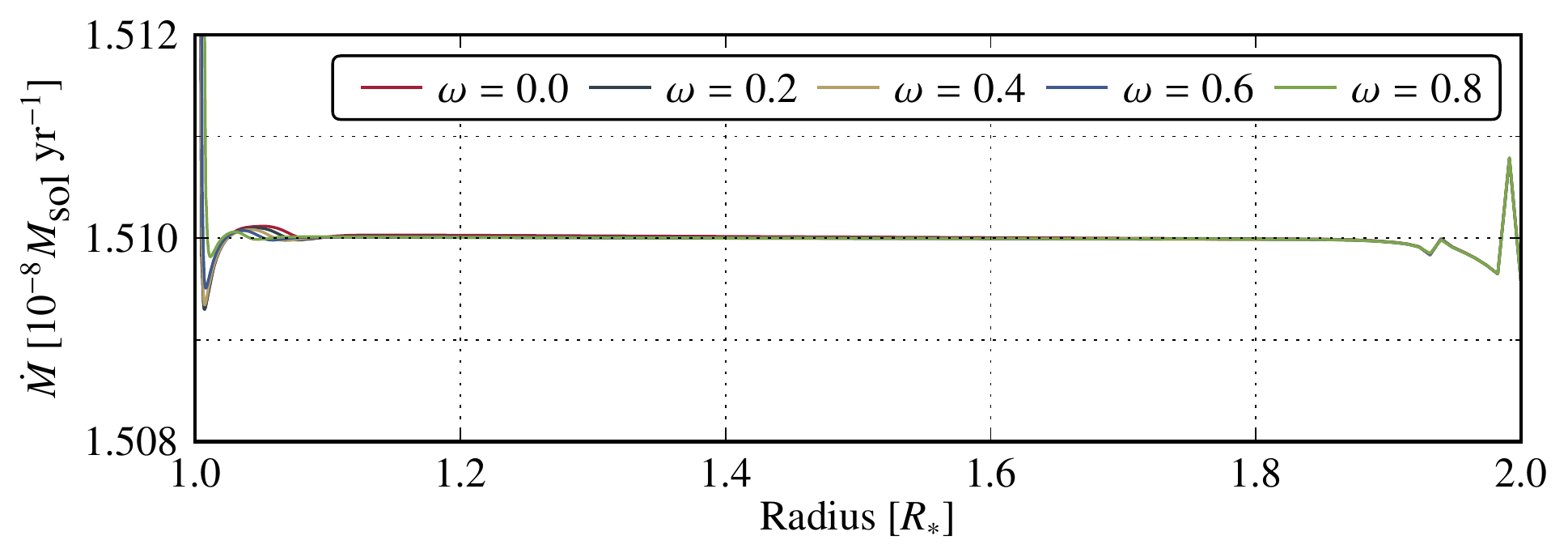}
    \caption{\label{fig:10_mdot}Mass accretion rate $\dot{M}=-2\pi r\Sigma u_r$ for $M_\ast=1.0M_\sun$ and five different stellar rotation rates
$\omega=\Omega_\ast/\Omega_\text{K}(R_\ast)$. The imposed value equals $\dot{M}=1.51\times10^{-8}M_\sun\ \text{yr}^{-1}$. Since $\dot{M}$ is constant in a stationary
state, this plot is also an indication that a good equilibrium has been reached.}
  \end{center}
\end{figure}

For a stationary state the continuity Eq.~(\ref{eq:cont_eq}) reduces to $\dot{M}=-2\pi r\Sigma u_r$, where $\dot{M}$ is the constant mass accretion rate that
represents the amount of mass flowing through an annulus at a given radius per time. Since we imposed a mass accretion rate of $\dot{M}=1.51\times10^{-8}M_\sun/\text{yr}$
at the outer boundary, we require $\dot{M}$ to attain this value throughout the disk if a steady state is reached. The mass accretion rate after a simulation time of
about $10\,000$ orbits at $r=R_\ast$ is shown as an example in Fig.~\ref{fig:10_mdot} for the $1.0\,M_\sun$ model. 
At first glance, the attained
equilibrium state looks very good and stable. The maximum deviation from the imposed value is only about $4\%$ for the heaviest white dwarf. The most constant $\dot{M}$ is reached for the models that
correspond to $\omega = 0.8$. Apart from the inner and outer boundary, where $\dot{M}$ matches the imposed value to an accuracy of far less than one percent, the agreement
between the simulation and the imposed value is perfect. This holds true for any of the three stellar masses. 
However, with decreasing stellar rotation rate and increasing stellar mass the mass accretion rate starts to deviate slightly from the constant value of $1.51\times10^{-8}M_\sun/\text{yr}$.  
The deviations near the outer boundary, however, are induced by the boundary conditions.

\begin{figure}[t]
  \begin{center}
    \includegraphics[scale=.45]{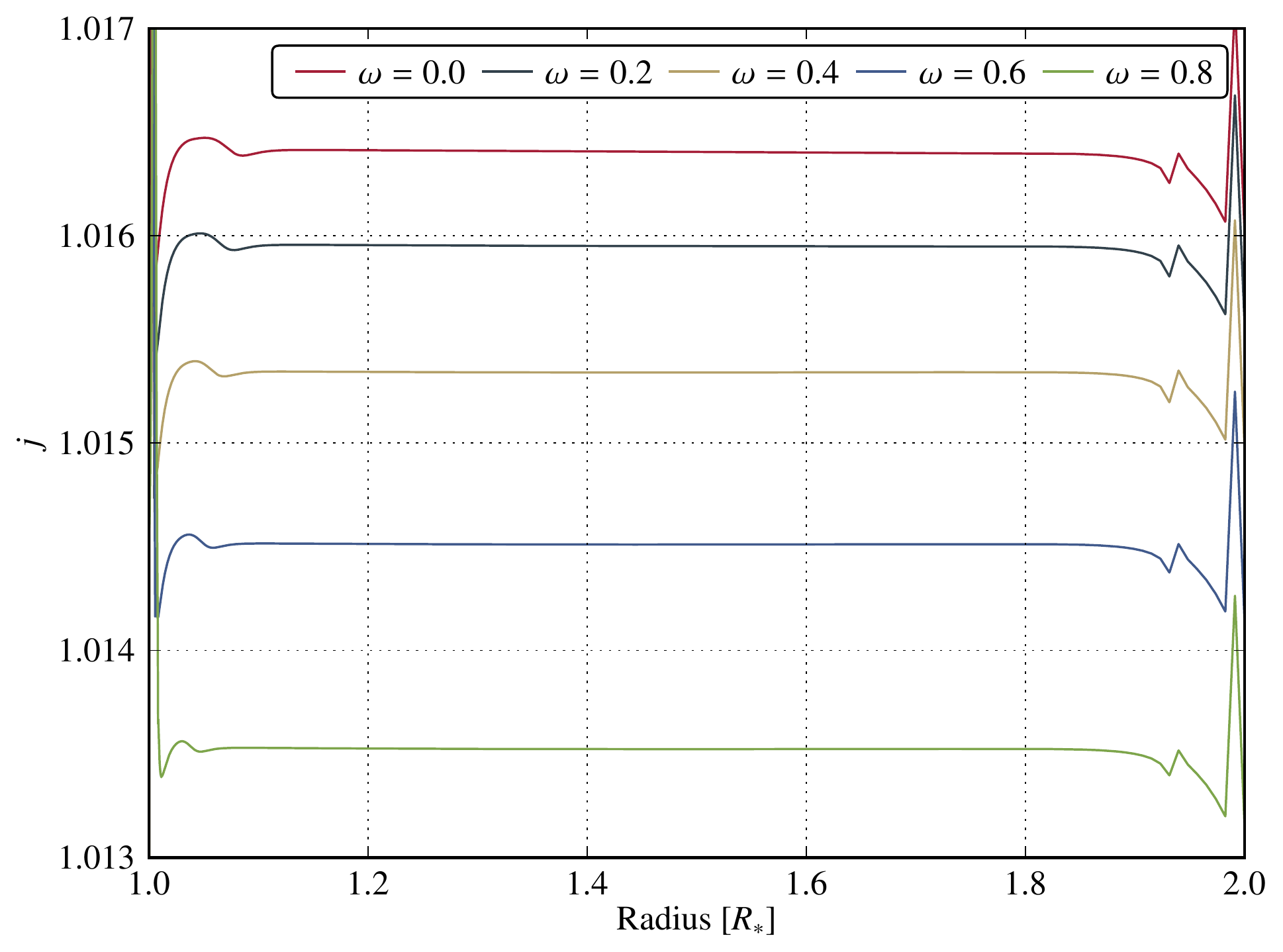}
    \caption{\label{fig:10_jdot}This plot shows the normalized angular momentum flux $j=\dot{J}/\dot{J}_\ast$, where $\dot{J}$ is given by Eq.~(\ref{eq:jdot}) and
$\dot{J}_\ast=\dot{M}R_\ast^2\Omega_\text{K}(R_\ast)$ is the advective angular momentum flux at the surface of the star. Displayed is a stellar mass of $1.0M_\sun$ and
five different stellar rotation rates in Keplerian units $\omega$. We note that $j$ should be constant in a stationary state.}
  \end{center}
\end{figure}

Another quantity that should attain a constant value in the steady state, is the angular momentum flux $\dot{J}$. It plays the role of an eigenvalue in the stationary
equations, whose value has to be determined while solving the set of equations \citep{1995ApJ...442..337P,1997MNRAS.285..239K}. In
our model, the total angular momentum flux is composed of the angular momentum carried in with the accreting material, and the angular momentum transported by shear
viscosity. Therefore, it is given as
\begin{equation}
	\dot{J} = \dot{M}r^2\Omega + 2\pi r^2\sigma_{r\varphi}~.\label{eq:jdot}
\end{equation}
Usually, the angular momentum (AM) flux is displayed as the normalized, dimensionless $j$, which is $\dot{J}$ from Eq.~(\ref{eq:jdot}) divided by the advective AM flux
at the surface of the star, $\dot{J}_\ast = \dot{M}R_\ast^2\Omega_\text{K}(R_\ast)$. This value is shown in Fig.~\ref{fig:10_jdot} for the $1M_\sun$ star. 
Again, good equilibrium states have been reached, since the deviations of $j$ are very small. We also note the same trend as for the mass accretion rate: with increasing stellar mass and decreasing stellar rotation rate the deviations gain in strength. Since $j$ is an eigenvalue in the stationary equations, it should be constant throughout the domain. For convenience, $\dot{M}$ is defined to be positive when mass is flowing to the center of the accretion disk and $\dot{J}$ is
positive for inward-moving angular momentum (see Eq.~\ref{eq:jdot}). Therefore, the net flux of angular momentum is directed inward and the absolute value is slightly
above the advected angular momentum at the stellar surface and equals the accreted angular momentum of the star. The mean values of the normalized total angular momentum
fluxes for $\omega=0.0$ and $\omega=0.8$ are given in Table \ref{tab:jdot}, along with the numerical value of the corresponding $\dot{J}_\ast$ in cgs-units. These values of $j$ correspond very nicely with those found by other numerical calculations, for example by \citet{1995ApJ...442..337P} who obtained a value of $j=1.01$ for a $0.6M_\sun$ mass white dwarf with a radius of $8.7\times10^{8}\,\text{cm}$ and a mass accretion rate $\dot{M}=10^{-8}M_\sun\,\text{yr}^{-1}$. Other deviations from the parameters used here are the usage of a different viscosity prescription and $\alpha=0.1$. We have run a test calculation with their set of parameters and were able to reproduce $j$ nearly perfectly.

The advected AM flux is pointing inward throughout the disk since the radial velocity $u_r$ only attains negative values, that is, matter is only moving to the center of the disk.
The transport of AM due to shear viscosity, on the other hand, changes its direction at the zero-gradient location, which is the radius where $\partial\Omega/\partial r =
0$ (and also the beginning of the BL, coming from the outside). Thus, in the disk, angular momentum is transported outward by the shear viscosity, while in the BL, it is
transported inward. The sum of $\dot{J}_\text{adv}$ and $\dot{J}_\text{visc}$ remains constant. This means that in the disk, where $\dot{J}_\text{visc}< 0$, the
advected angular momentum flux must be greater than the total flux, $\dot{J}_\text{adv}>\dot{J}$. In the BL, however, the viscous AM flux becomes positive and
therefore $\dot{J}_\text{adv}<\dot{J}$. At the inner boundary, $\dot{J}_\text{adv}$ drops to nearly zero because $u_r$ drops to nearly zero. Hence, $\dot{J}_\text{visc}$ must be
approximately equal to $\dot{J}$ although there is only very weak shearing. Here, $\nu\Sigma$ becomes very large, since the radial velocity is very small
and the temperature is very high, which produces a high viscosity. If we compare the models with different stellar masses, we find that $\dot{J}_\ast$ gets smaller with
increasing stellar mass (see Table \ref{tab:jdot}). Two points affect the absolute value of $\dot{J}_\ast$ for different stellar masses. On the one hand, there is the
mass-radius-relation of white dwarfs, meaning that with increasing mass, the stellar radius gets smaller. On the other hand, with increasing stellar mass, the
Keplerian angular velocity also increases. Both effects together cause $\dot{J}_\ast$ to decrease only weakly with increasing stellar mass. The
values of $j$ also decrease with increasing stellar mass. While $\dot{J}_\text{adv}$ gets slightly bigger with increasing stellar mass ($\dot{M}$ is equal for all
models and equals the imposed value while the angular velocity increases), $\dot{J}_\text{visc}$ also has to be larger to
yield a smaller value of $j$. The reason for the increasing AM transport by viscosity is that the disk has a higher surface density and midplane temperature and thus a
greater viscosity. We also observe a clear trend concerning the various values of $j$ for different stellar rotation rates $\omega$ and constant stellar mass. With increasing stellar rotation rate $\Omega_\ast$, $j$ clearly decreases, as it also does with increasing stellar mass and constant rotation rate. Both trends were also found by \citet{1995ApJ...442..337P}.
From Eq.~(\ref{eq:jdot}) we deduce that in the disk, $\dot{J}_\text{adv}$ must be nearly the same for every $\omega$. In the BL, by contrast,
it increases with higher stellar rotation rate since $\Omega$ does not drop as much. It is harder to make a reliable point concerning $\dot{J}_\text{visc}$. In the
disk, however, both the midplane temperature and the density are slightly greater for increasing $\omega$ (see Figs.~\ref{fig:10_Sigma} and \ref{fig:10_Tc}), yielding a
higher viscosity and hence a greater $\dot{J}_\text{visc}$. This explains the smaller values of $j$ in the disk. In the BL, however, the situation is not as clear, since the surface density increases with $\omega$ while the midplane temperature decreases.

\begin{table}
\caption{\label{tab:jdot}Normalized angular momentum flux $j$ for three different stellar masses and two different stellar rotation rates. The
		numbers shown are a constant fit to the results.}
	\begin{center}
		\begin{tabular}{c c c c}
		\hline\hline
		model & $j(\omega=0.0)$ & $j(\omega=0.8)$ & $\dot{J}_\ast$ $[\text{cm}^2\ \text{g}\ \text{s}^{-2}]$\\
		\hline
		$M_\ast = 0.8 M_\sun$ & $1.0187$ & $1.0155$ & $2.6088\times10^{35}$ \\
		$M_\ast = 1.0 M_\sun$ & $1.0164$ & $1.0135$ & $2.5846\times10^{35}$ \\
		$M_\ast = 1.2 M_\sun$ & $1.0140$ & $1.0119$ & $2.3915\times10^{35}$	\\
		\hline
		\end{tabular}
	\end{center}
\end{table}

\subsection{Comparison with other BL models and equatorial radius increase}\label{sec:comparison_pn95}

In this section we put our simulations in context with other work, that has been done in the field of boundary layers. As has been mentioned before, our results are closely related to those presented in \citet{1995ApJ...442..337P} (hereafter PN95), at least qualitatively. However, the approach we took is quite different from this work. While \citeauthor{1995ApJ...442..337P} have used the time-independent equations and performed stationary calculations, we have also propagated the physical quantities in time. Therefore, we are able to study time-dependent phenomena as well and we are not limited by the prerequisite that a stationary state exists for the given choice of parameters. Also, as we have seen in the last section, the BL is in permanent motion, even though most physical quantities do not change perceptibly. We have seen this by analysing the constant parameters $\dot{M}$ and $j$. Other differences involve the treatment of the radiation in the equations. While the authors of the quoted publication have taken care of the radiation field by adding an expression to the gas pressure, hence defining a total pressure, we have explicitly added the radiation force term in the momentum equation and a special pressure work term in the energy equation, both in the flux limited diffusion approximation. Although both approaches are identical in the optically thick limit, this is not necessarily the case for the optically thin and transition regions.

In order to compare our results quantitatively with the computations of \citet{1995ApJ...442..337P}, we ran a set of simulations with the same parameters, the basis of which is their standard model. It is composed of a $0.6 M_\sun$ white dwarf with a radius of $8.7\times 10^{8}\,\text{cm}$ and a mass accretion rate of $\dot{M}=10^{-8}M_\sun/\text{yr}$. The white dwarf does not rotate and they used a special viscosity prescription (see Eq.~\ref{eq:pn95_viscosity}) and a $\alpha$-parameter of $0.1$. For the standard model, we found a dynamical BL width of $\Delta r=1.017R_\ast$, which is exactly the same as in PN95, and an AM flux of $j=1.004$, as opposed to $j_\text{PN95}=1.00994$. The dynamical quantities, viz. $\Omega$ and $u_r$, match the calculations of PN95 very closely ($\sim 1\%$). Only the peak value of the infall velocity deviates by about $10\%$. The peak position, on the other hand, is accurate to about $0.4\%$. The good agreement is also reflected in the thermodynamical quantities, namely the disk and the surface temperatures $T_\text{c}$ and $T_\text{eff}$. The second has a peak value in the BL of $227\,000\,\text{K}$ and about $54\,000\,\text{K}$ in the disk. Both temperatures agree to approximately one percent with the results of PN95. While the peak position of the midplane temperature $T_\text{c}$ accurately matches that of PN95, our $T_\text{c}$ as a whole is about $30\%$ hotter. It is, however, $T_\text{eff}$ that determines the emergent spectra of the BL and the disk, and since we find a very good match here, we conclude that our simulations are in very good agreement with the results of PN95.

Starting from the standard model described above, we performed the same $\Omega_\ast$ parameter study as in \citet{1995ApJ...442..337P}. We have done this in particular to state an important point that has not been taken into account in the simulations shown above. Since the white dwarf flattens out considerably with increasing $\Omega_\ast$, in principal we have to consider an equatorial radius increase. The radius of a rotating white dwarf in turn has to be calculated from stellar structure simulations. However, one finds \citep[see e.g.][]{1986ApJS...61..479H} that for moderately rotating white dwarfs ($\omega\lesssim 0.8$), $R_\ast$ increases at most by a factor of $\sim1.4$. The change in stellar radius affects the effective temperature of the BL and the disk through Eq.~(\ref{eq:acc_en}), yielding slightly smaller values. This is in agreement with the standard solution for accretion disks, where the radius enters the effective temperature to the power $-3/4$. Because of the non-linear variation of gravity, the width of the BL also changes somewhat. Owing to the minor effect of the small radius increase and since the majority of white dwarfs are supposed to be slow rotators \citep{1998ApJ...505..339L,2012MmSAI..83..539S}, we have neglected this effect in the simulations above. Nevertheless, we want to illustrate the equatorial radius increase on the basis of the fastest rotating white dwarf in the aforementioned $\Omega_\ast$ study, which has a rotation rate in terms of the fraction of the breakup rotation rate of $\omega=0.86$.

The width of the dynamical BL changes from $\Delta r\approx 1.018R_\ast$ for the model with equatorial radius increase (M1, $R_\ast\approx 1.25\times 10^{9}\,\text{cm}$) to $\Delta r\approx 1.017R_\ast$ for the model without an increase in radius (M2), i.e. $R_\ast=8.7\times 10^{8}\,\text{cm}$. This is a variation of about one per mill. The difference concerning the normalized AM flux $j$ between both models is even smaller, where $j_\text{M1} = 1.0038$ and $j_\text{M2} = 1.0033$. The agreement of the angular velocities is very good, as can be seen from the width of the BL. The radial velocities of both models do match each other very closely, as well. However, the absolute peak value of M2 is about $15\%$ greater than that of M1. If instead we compare the Mach numbers, the deviation shrinks to about $1\%$, but the differing radii manifest themselves more obviously in the disk and effective temperature of the models. Thus, the disk temperature $T_\text{c}$ of model M2 is about $30\%$ higher throughout the whole domain which comes to approximately $40\,000\,\text{K}$. This is due to the greater gravitational pull for smaller radii. We also find the same $30\%$ deviation in the surface temperature $T_\text{eff}$, although here it accounts for only $13\,000\,\text{K}$ because of the lower temperature regime of the surface temperature. Apart from this vertical shift, both graphs are identical, meaning that the peak is at the same location and the thermal BL has the same radial extent. Model M1 peaks at $\sim65\,000\,\text{K}$ in the BL.

\citet{1995MNRAS.275.1093G} used a time-dependent spectral numerical code and performed one-dimensional calculations of the BL that are closely related to ours. The aim of their simulations was to investigate the influence of the thermal inner boundary condition, which can lead to both cool and hot BLs, depending on whether the temperature is held fixed at the inner boundary or the flux of the star is fed into the computational domain. We have performed a simulation for the case of flux BC with the same parameter choice as in \citet{1995MNRAS.275.1093G}, their model 3, and were able to reproduce the results very closely.

\subsection{Simulations with larger domain}

For the Navier-Stokes equations it is still uncertain whether the same solution is reached for a small, bounded domain as for an unbounded domain. Of course, we cannot simulate our problem on an unbounded domain. We can, however, prove that the solution we presented above does not depend on the choice of the domain. In order to do this, we ran a test simulation, where we enlarged the computational domain by a factor of $10$, so that it reaches from $r=1R_\ast$ up to $r=10R_\ast$. Except for the number of grid cells, all parameters were chosen according to the $0.8M_\sun, 0.4\Omega_\text{K}$ model. We find a nearly perfect agreement between the models with the small and the large domain. Differences are, apart from the boundaries, nearly imperceptible, except for the disk temperature, where the larger domain overestimates the peak temperature in the BL by approximately $5\%$. This is, however, due to the lower resolution of the model in this region.

Finally, to show the agreement of our models with the standard solution for accretion disks by \citet{1973A&A....24..337S}, we compare our results with the analytically derived formulae for thin accretion disks. Since these equations are supposed to be a good approximation in the disk, we used the model with the large domain for the comparison, so that we could also compare both solutions in the disk over a farther region. The equations of a Shakura-Sunyaev-type solution can, for example, be found in \citet{frank2002accretion}. In these equations, we have modified the typical factor $[1-(R_\ast/R)^{1/2}]$ by a factor of the form $[1-j(R_\ast/R)^{1/2}]$, where $j$ is the normalized angular momentum flux. The solution cuts off not at $R_\ast$, but a little farther outside ($\sim 1.02R_\ast$). We find that the S-S-type solution provides a very good approximation of the physical variables in the disk. In the BL the standard solution is insufficient. We also find that the surface temperature of our simulation peaks at $r=1.4092R_\ast$ in the disk (there is also a far more distinct peak in the BL). The S-S standard solution for the surface temperature reads
\begin{equation}
	T(r) = \left\{\frac{3GM_\ast\dot{M}}{8\pi r^3\sigma_\text{SB}}\left[1-j\left(\frac{R_\ast}{r}\right)^{1/2}\right]\right\}^{1/4}~.\label{eq:teff_standard}
\end{equation}
If we plug $j=1.0175$ into Eq.~(\ref{eq:teff_standard}), which is the normalized angular momentum flux for this simulation, we find that the above function $T(r)$ has a maximum at $r\approx1.4093$. Thus, our simulation is in perfect agreement with the theoretically derived formula. The temperature in Eq.~(\ref{eq:teff_standard}) peaks at a value of $T_\text{max}\approx0.475T_\ast$, where $T_\ast$ is defined by $T_\ast=\left(\frac{3GM_\ast\dot{M}}{8\pi R_\ast^3\sigma_\text{SB}}\right)^{1/4}$. According to (\ref{eq:teff_standard}), for our model this means that the temperature peaks at $T_\text{max}=74\,406\,\text{K}$. The simulation shows a temperature of $T=74\,335\,\text{K}$ of the peak of the surface temperature in the disk. Again, this is in perfect agreement with the theoretical value. Having matched the other quantities of our simulations against the S-S standard solution as well, we come to the conclusion that outside the BL, our simulations are perfectly described by the standard solution.

\section{Summary and conclusion}

We have presented new models of the structure of the boundary layer around white dwarfs in compact binary star systems.
One-dimensional, time dependent radial models have been constructed which include radiative diffusion in the radial direction, vertical
cooling from the disk surfaces, and radiative pressure effects.

For a fixed mass accretion rate of $\dot{M} = 1.51 \times 10^{-8} M_\odot$/yr, which corresponds to systems in outburst,
we have analyzed the BL for different masses and rotation rates of the white dwarf.

The strong shear in the BL region leads to an enormous energy release and to surface temperatures of a few hundred thousand Kelvin.
For a non-rotating white dwarf (with $1 M_\odot$) the maximum temperature is about $500\,000\,\text{K}$, while for
a star rotating with $\omega =0.8$ of the break-up velocity the maximum is about $200\,000\,\text{K}$.
Hence, knowledge of the white dwarf mass, for example through a dynamical mass estimate of the binary star, and of the mass accretion rate,
allows an estimate of the stellar rotation rate through the observed peak temperatures.
Radial diffusion of energy leads to a more extended thermal BL with a width of typically 0.02 to 0.05 $R_*$.
The models for slow rotation showed a tendency for instability due to the very high temperatures, small vertical thicknesses,
and resulting low optical depths.

For the viscosity we use the $\alpha$-parametrisation with $\alpha=0.01$. For this value, the radial velocity remains subsonic throughout
with maximum radial Mach numbers of 0.35 for $\omega =0$ and 0.18 for $\omega =0.8$. A higher value of $\alpha=0.1$ left the disk structure
unchanged and Mach numbers close to unity within the BL.
 
Varying the stellar mass leads to hotter BL for larger masses (and smaller radii) and cooler BL for smaller masses.
Hence, when trying to infer stellar parameter through an analysis of the BL radiation one is faced with an ambiguity that
models with different combinations of $R_*$ and $\omega_*$ may yield similar peak temperatures. There is the indication, however, that
the width of the thermal BL is different in these cases, such that the model spectra will lead to different results.
In this paper we did not calculate synthetic, theoretical spectra for our numerical models, but leave that for a future paper.

The validity of our simulations has been demonstrated convincingly by comparing the results with related calculations and with the standard solution for thin accretion disks by \citet{1973A&A....24..337S}. We found very good agreement with our results, both in the disk where the standard solutions holds true, and in the BL, where we were able to reproduce basic features shown in other works and match their results to an accuracy of about one percent for the dynamical and observed quantities. We also showed that our results are independent of the computational domain and the resolution. Simulations that have the same parameters, except for the simulation area, do match each other perfectly.

An equatorial radius increase due to a flattening of a fast rotating white dwarf in our simulations showed the following results. Even for the unlikely case of a rotation rate of $86\%$ of the breakup rotation rate, the variation of the physical quantities that differ most is in the range of $30\%$, according to whether we take the flattening into account or not. The width of the thermal BL, which is also important for the emergent spectra, does not change perceptibly, however. For a $0.6M_\sun$ white dwarf, the shift in the effective temperature is of the order of $10\,000\,\text{K}$.

Analysing the data of our simulations, we also found that the consideration of radiation energy is indeed necessary in our models. We see this from the radiation pressure $P_\text{rad}=aT^4/3$, which becomes comparable to the thermal pressure $P$ in the BL and even exceeds it by a factor of the order of unity for a small radial extent. If $P_\text{rad}$ is not taken into account in the simulations, the effective temperature peaks at far higher temperatures and the width of the thermal BL is distinctly smaller because of the lack of the widening effect of $P_\text{rad}$.

Owing to the slim disk approximation, our models do not allow us to answer the question of how the material settles onto
the central white dwarf. This question can only be answered by two dimensional $r$-$z$ simulations, similar to those
by \citet{2009ApJ...702.1536B} but with radiative transport \citep{1991A&A...247...95K}. 
The strong shear in the BL can lead to unstable behavior when considering the $\varphi$-direction, as described by 
\citet{2012ApJ...760...22B} for isothermal disks. This behavior could not be found in our simulations as they are purely
radial. The next step would be to extend these to two-dimensional $r$-$\varphi$ disks.

\begin{acknowledgements}
      M. Hertfelder received financial support from the German National Academic Foundation (Studienstiftung des deutschen Volkes).
      The work  of V. Suleimanov is supported by the German Research Foundation (DFG) grant SFB/Transregio 7 "Gravitational Wave Astronomy".
      We also thank the referee for his constructive comments which helped to improve this paper.
\end{acknowledgements}

\bibliographystyle{aa}
\bibliography{references}
\end{document}